\documentclass[a4paper]{article}

\usepackage[english]{babel}
\usepackage[utf8x]{inputenc}
\usepackage[T1]{fontenc}

\usepackage[a4paper,top=3cm,bottom=2cm,left=3cm,right=3cm,marginparwidth=1.75cm]{geometry}

\usepackage{amsmath, amssymb}
\usepackage{textcomp}
\usepackage{gensymb}
\usepackage{amsthm}
\usepackage{graphicx}
\usepackage{caption}
\usepackage{subcaption}
\usepackage{adjustbox}
\usepackage{booktabs}
\usepackage{pdfpages}
\usepackage{placeins}

\usepackage[colorlinks=true, allcolors=blue]{hyperref}
\usepackage[numbers,sort&compress]{natbib}

\usepackage{setspace}
\title{Enhanced contact flexibility from nanoparticles in capillary suspensions}
\author{Lingyue Liu, Jens Allard$^{\dag}$, Erin Koos$^{\ast}$}
 \date{ \small
 {KU Leuven, Department of Chemical Engineering, Celestijnenlaan 200J, 3001 Leuven, Belgium} \\
$^{\dag}$ {Current address: Robert Bosch Produktie N.V., 3300 Tienen, Belgium} \\
$^{\ast}$ E-mail: erin.koos@kuleuven.be \\}

\begin{document}

\maketitle

\begin{abstract}
\emph{Hypothesis:} 
Sample-spanning particle networks are used to induce structure and a yield stress, necessary for 3D printing of porous ceramics and paints. In capillary suspensions, a small quantity of immiscible secondary fluid is incorporated into a suspension. By further adding nanoparticles with a range of hydrophobicities, the structure of the bridges and microparticle-microparticle contacts should be modified, resulting in a tunable yield stress and shear moduli. Moreover, the compressibility of these samples, important in many processing and application steps, should be sensitive to these changes. 

\emph{Experiment:}
The nanoparticle hydrophobicity was altered and their position relative to the microparticles and the bridges was examined using confocal microscopy where the correlation between bridge size and network structure was observed. A step-wise uniaxial compression test on the confocal was conducted to monitor the microparticle movement and structural changes between capillary suspension networks with and without nanoparticles.

\emph{Findings:}
Our observation suggests that nanoparticles induce the formation of thin liquid films on the surface of the microparticles, mitigating contact line pinning and promoting internal liquid exchange. Additionally, nanoparticles at microparticle contact regions further diminish Hertzian contact, enhancing the capacity for rearrangement. These effects enhance microparticle movement, narrowing the bridge size distribution. 
\end{abstract}

\section{Introduction}

Nanoparticles have been used in systems such as dense suspensions~\cite{Yang2017}, dry granulates~\cite{Yang2005, Kim2022} and wet granulates~\cite{Friedrich2010}, to modify their rheological properties. For example, the dynamic viscosity is reduced in dense suspensions with a bimodal size distribution~\cite{Gondret1997, Probstein1994}. The viscosity is reduced to the greatest extent for more extreme size ratios and intermediate fractions of small spheres ranging from 20--50~\%~\cite{Chong1971, Greenwood1997}. Small fractions of nanoparticles can even have a negative effect: increasing the aggregation of particles due to the depletion interaction for either uncharged or 1~\% weakly adsorbing nanoparticles~\cite{McKee2012}. In granular systems, nanoparticle excipients can increase powder flowability by either dry coating the particles~\cite{Yang2005, Kim2022} or even by dry coprocessing~\cite{Kim2023}. The nanoparticles act like flow aids, reducing the Hamaker constant and cohesive forces between microparticles to enhance sliding. Enhancing contact flexibility in systems with an existing particle network remains challenging, however. This is especially the case in systems, such as wet granular media and capillary suspensions where the nanoparticles do not simply interact with the bulk phase and the microparticles, but also liquid bridges. 

Capillary suspensions, formed by adding a small amount of an immiscible fluid to a normal suspension, are distinct from either wet granular media and traditional suspensions~\cite{Koos2011}. The solid particles within the capillary suspensions are held together via the capillary attraction provided by the secondary liquid menisci, creating a sample-spanning network throughout the bulk liquid. Based on the three-phase contact angle, capillary suspensions can be classified as either in the pendular state or the capillary state~\cite{Koos2014, Bossler2016, Bindgen2020}. Pendular state suspensions occur when the secondary liquid has a preferential wetting angle smaller than 90$^{\circ}$, forming concave menisci, while the capillary state suspensions are formed when the secondary liquid has a non-preferential wetting angle greater than 90$^{\circ}$, forming convex menisci. In both cases, the formation of a sample-spanning network significantly increases the viscosity, inducing a gel-like rheological behavior when compared to binary suspensions without the secondary liquid~\cite{Koos2014}. Capillary suspensions have been applied to various applications based on their tunable rheological properties, such as crack-free coatings, low-fat spreadable chocolate, printable electronics, and as a precursor for the fabrication of porous ceramics~\cite{Dittmann2016, Fischer2021a, Hoffmann2014, Dittmann2013, Wollgarten2016}.

When nanoparticles are added to capillary suspensions, they are expected to influence several aspects of the preparation and structure. The presence of nanoparticles impacts not only the characteristics of the base fluid but also the nanoparticle-microparticle interaction~\cite{Aksoy2023, Park2019, Dittmann2016}. For instance, the increased shear and interfacial viscosity can influence the homogenization of secondary liquid inside the bulk liquid and would, therefore, influence the bulk mechanical properties such as the yield stress~\cite{Bossler2017}. When added with the secondary fluid, Wei{\ss} et al.~\cite{Weis2020} found that the nanoparticles were predominantly located in the regions between the ceramic particles and served to form dense sintering bridges. The compressive strength was identical for both alumina and aluminosilicate microparticles, indicating that the strength is dominated by the silica nanoparticles in the bridges~\cite{Weis2020}.  The localization of nanoparticles in the bridges was confirmed by Park et al.~\cite{Park2019}  who used Cryo-SEM to image fatty acid coated styrene-butadiene rubber (SBR) nanoparticles between adjacent mildly hydrophilic graphite microparticles~\cite{Park2019}. In both of these cases, the nanoparticles had favorable wettability with the secondary liquid. The structure and location of nanoparticles with intermediate or unfavorable wetting remains unknown. Further, the influence of these nanoparticles on the rheological response has also not yet been investigated.

When introducing freely moving nanoparticles into capillary suspension systems, the nanoparticles also fundamentally alter the secondary liquid response. They can influence the contact angle of secondary liquid on microparticles, dynamic behavior of bridges stretching and breaking and friction contribution of microparticle movement during shearing~\cite{Bossler2016, Bossler2017, Bossler2018, Allard2022, Tevet2011, Dai2016, Dittmann2016, Park2019}. When nanoparticles are adsorbed irreversibly and act as microparticle surface roughness, the increase in the degree of roughness (nanoparticle size) results in corresponding transition between wetting regimes~\cite{Allard2022}. As shown for wet granular materials, limited bridging fluid results in liquid that is confined between large asperities in the asperity regime~\cite{Halsey1998}. With increasing fractions of the secondary liquid, the system passes through a roughness regime, were many small bridges form and the total interaction force increases proportionally with the amount of secondary liquid, and then the spherical regime where large, pendular bridges form between the particles~\cite{Halsey1998}. As roughness increases in the capillary suspension system, the amount of trapped liquid increases which induces a reduction in capillary bridge size and number leading to a decrease capillary forces~\cite{Allard2022}. Even when the effective bridge size is equalized, however, the roughness can also act as mechanical barrier, thereby increasing the length of linear viscoelastic regime and forming structures with fewer contacts and less clustering~\cite{Allard2022}. The influence of freely moving nanoparticles on the secondary liquid response, however, remains unknown. 

In the present work, we apply fluorescently labeled nanoparticles to study their energetically favorable positions in correspondence to their hydrophobicities. We also apply step-wise uniaxial compression experiments on capillary (nano)suspensions and reveal the changes in liquid bridge size distribution and gel structure via 3D tracking. The influence of the mechanical response to compression as well as in shear is used to elucidate the interactions between particles both with and without added nanoparticles.

\section{Materials and methods}

\subsection{Solid phases}

Silica microparticles (SOLAD non-porous PNPP10.0NAR Glantreo) with an average diameter of 10 $\mathrm{\mu}$m and a polydispersity of under 5 \% were used as the solid phase of the capillary suspensions. Silica nanoparticles were synthesized via the St\"{o}ber process. The density of silica particles, 2.15 $\pm$ 0.11 g/ml, was determined via buoyancy measurement both in air and in ethanol (Sartorius Density Determination Kit YDK01). Both silica micro- and nanoparticles were selectively, fluorescently labeled using rhodamine B isothiocyanate (RBITC, Sigma-Aldrich). When imaging the nanoparticle migration mechanism, a modified core-shell technique was applied to avoid the hydrophobic effect that excessive dye molecules might cause due to the limited surface area of nanoparticles. When imaging microparticle structures, silica microparticles were tagged via a modified St\"{o}ber synthesis, which covalently bonded the dye molecules to the particle surfaces after which another St\"{o}ber process was carried out to coat the particles with a pristine silica layer.

To produce fluorescent nanoparticles without affecting the intrinsic characteristics of their surfaces, a modified St\"{o}ber core-shell structure was used. First, to produce the dye solution, 10~mg rhodamine B isothiocyanate powder (RBITC, Sigma-Aldrich), 10~$\mathrm{\mu}$l 3-Aminopropyl)triethoxysilan (APTES, Sigma-Aldrich) and 15~ml anhydrous ethanol (99.8~\%, Fischer Chemical) were mixed together and stirred overnight at room temperature (20.5 $\pm$ 1~{\textdegree}C) to ensure a complete conjugation. Afterwards, 3~ml of the dye solution was mixed with 0.894~ml tetraethyl orthosilicate (TEOS, 98\%, Acros Organics), 17.6~ml anhydrous ethanol (99.8\%, Fischer Chemical) and 1~ml ammonium hydroxide solution (28-30\%, Sigma-Aldrich) then stirred with 300~rpm at room temperature (20.5 $\pm$ 1~{\textdegree}C) for 18~h. During this step, the ethoxy groups of the conjungated RhB-APTES were hydrolyzed and form silanol groups, on which the silanol groups of the hydrolyzed TEOS was able to condensate further and crosslink, thus, suspensions with fluorescent silica cores of 90 $\pm$ 6~nm were created, tested using a commercial 3D dynamic light scattering device (3D DLS, LS Instruments). The produced suspensions were centrifuged at 16~{\textdegree}C using an Avanti J-30I Centrifuge (Beckman) at 9000~rpm for 20~min. Three rinsing cycles with ethanol were used before the particles were redispersed in anhydrous ethanol. The nanoparticle suspension weight percentage was determined using Thermogravimetric analysis(TGA)$-$Q500 (Waters L.L.C.) and diluted to 0.1~vol\% with anhydrous ethanol. Afterwards, 3~vol\% TEOS and 4~vol\% ammonia were added into the diluted nanoparticle core suspensions and stirred with 300~rpm at room temperature (20.5 $\pm$ 1~{\textdegree}C) for 18~h. During this step, TEOS was hydrolyzed in the presence of ethanol and ammonia and further condensate onto the dyed cores and formed undyed shells with a thickness of 30~nm, resulting in a final particle size of 150 $\pm$ 20~nm. 

The washing process was repeated and the acquired samples were rediluted to 0.5~vol\% with anhydrous ethanol. Various amounts of hexamethyldisilazane (HMDS, Acros Organics B.V.B.A.) were added into the anhydrous ethanol suspension with 150~nm core-shelled particles and stirred with 300~rpm at room temperature (20.5 $\pm$ 1~{\textdegree}C) for 18~h. The hydrophobization of silica nanoparticles is caused by HMDS molecules chemically bonding its Si atom to the oxygen atom from the silanol functional groups (Si-OH) on nanoparticle surfaces~\cite{Suratwala2003}. The ratio of applied hexamethyldisilazane ($\mathrm{\mu}$l) to silica nanoparticle (mg) in the suspension is abbreviated as HSR, with HSR0 representing untreated hydrophilic nanoparticles. Finally, the samples were again rinsed 2 times using ethanol and 3 times using water, and the final products were redispersed in water. The index of refraction of the dyed core-shell nanoparticles was $n=1.455$. The hydrophobized nanoparticles are characterized by testing their zeta-potential using the NanoBrook omni (Brookhaven Instruments). The resulting $\zeta$-potentials for each amount of HMDS are listen in Table~\ref{tab:zeta-potential}.
\begin{table}[!tb]
  \centering
  \caption{Zeta-potential measurements of nanofluids with different hydrophobicities where HSR is the ratio of HMDS to silica ($\mathrm{\mu}l$/mg).}
  \begin{tabular}{l c c}
    \toprule
    \textbf{Nanoparticle} & \textbf{$\zeta$-potential [mV]} \\
    \midrule
    undyed & -45.13 $\pm$ 1.48 \\
    HSR0 & -44.93 $\pm$ 0.09 \\
    HSR1 & -42.32 $\pm$ 0.92 \\
    HSR2 & -41.62 $\pm$ 0.7\\
    HSR4 & -35.8 $\pm$ 0.22 \\
    \bottomrule
  \end{tabular}
  \label{tab:zeta-potential}
\end{table}

\subsection{Capillary (nano)suspension preparation}

The bulk phase of the capillary suspensions, a mixture of 83.8 vol\% 1,2-cyclohexane dicarboxylic acid diisononyl ester (Hexamoll DINCH, BASF) and 16.2~vol\% n-dodecane ($>$ 99 \%, Sigma-Aldrich), has a refractive index of $n=1.455$, matching that of the silica microparticles. The density of this mixture is 0.91~g/ml. The secondary liquid is a mixture of 50~vol\% glycerol ($>$ 99.5 \%, Carl Roth) and 50 vol\% ultrapure water (Arium 611DI, Sartorius Stedim Biotech) with a $n=1.4$ and a density of 1.12~g/ml. The slight mismatch of the refractive index does not negatively influence the quality of the confocal images for the small amounts of secondary fluid and prevents problems during dispersion caused by elevated viscosity of index-matched aqueous glycerol. The samples with added nanoparticles were prepared at a volume fraction of $\phi_{NP} = 1$~vol\%.

To prepare the capillary suspensions, 10.1~$\mathrm{\mu}$l of the secondary liquid (either pure aqueous glycerol or the nanosuspension with $\phi_\text{sec}$ = 1 vol$\%$), which has been dyed with 1~vol\% PromoFluor 488 premium carboxylic acid (PromoCell GmbH), was added into 800~$\mathrm{\mu}$l bulk liquid. The interfacial tension between the aqueous glycerol and the Hexamoll DINCH/n-dodecane mixture is 25.5 $\pm$ 0.7 mN/m. The interfacial tension was measured with a drop tensiometer (Attension, Biolin Scientific) at room temperature (20.5 $\pm$ 1~{\textdegree}C) at a frame rate of 0.05 FPS and a duration of 30 min, the value was constant throughout the experiment period. The liquids were dispersed using a 2-step mixing procedure with an ultrasonic horn of 3.175~mm diameter (Digital Sonifier model 450, Branson Ultrasonics corporation). The fluids were first dispersed at a 10~\% amplitude for 10~s and 30~\% amplitude for 30~s, after which 435~mg of silica microparticles ($\phi_{MP}$=20~vol\%) was added into the sample. After the particles were added, the sample was again mixed using the ultrasound horn at 10~\% amplitude for 20~s.

\subsection{Rheological measurements}

A stress-controlled rheometer MCR702 (Anton Paar) with an 8 mm top plate and a 50 mm bottom plate geometry was used to perform oscillatory measurements. A double motor setup was used for enhanced sensitivity~\cite{MCR702}. All amplitude sweeps were conducted at an angular frequency of 10 rad/s. Frequency sweeps with a shear strain of 0.005~\% ensured the frequency independence of capillary (nano)suspensions between 0.1 rad/s and 10 rad/s.

The sample was loaded with a spatula. The top plate was lowered to a gap of 3 mm, after which the compression is recorded. The top plate was lowered to the trim position of 1.025 mm, 25~$\mathrm{\mu}$m above the measuring position, with an uniaxial velocity of 19.75~$\mathrm{\mu}$m/s. This allows the extra sample at the edges to be scraped prior to other measurements. Afterwards, the top plate is lowered to the  measurement gap of 1 mm. The normal force, i.e. the thrust on the plate, was recorded during both the gap lowering and oscillating experiments. Unless stated otherwise, the rheological properties are the average of at least three measurements with a normal force at the trim position of 33 $\pm$ 3~mN.

\subsection{Confocal microscopy and compression}\label{ssec:shellcell}

Capillary (nano)suspensions were loaded using a spatula onto a cover glass of dimension 22 x 40~mm (Menzel-Gläser) with a thickness of 130~$\mathrm{\mu}$m. Then, the loaded sample is sheared by a doctor blade at 350~$\mathrm{\mu}$m height, and subsequently trimmed into a 1 $\times$ 1 cm square at the center of the cover glass. This procedure ensured a reproducible starting condition for all compression experiments. The samples were then transferred onto the RheOptiCAD shear cell (CAD Instruments, Naucelle, France) attached to the Leica TCS SP8 inverted confocal laser scanning microscope (Leica Microsystems GmbH, Wetzlar, Germany). A glycerol immersion objective with 63x magnification and a numerical aperture of 1.3 was used to image the particle structures. Solid-state lasers with wavelengths of 488 nm and 552 nm were used to excite PromoFluor-488 (liquid channel) and rhodamine B isothiocyanate (particle channel), respectively. 

Due to the yield stress in capillary suspensions~\cite{Koos2011, Allard2022}, the trimming procedure with the doctor blade did not result in perfectly smooth upper free surfaces. To find the starting position for compression, the highest point of the network was first located using the confocal microscope. The top cover glass of the shear cell (24 $\times$ 60~mm) was lowered until contact between the top glass and the top layer of microparticles was observed. The compression test consisted of 20 steps of 3~$\mathrm{\mu}$m using the vertical stepper motor of the shear cell. This resolution was chosen to be smaller than microparticle radius so that the particle movement could be tracked precisely. Between each step, a 3D image stack consisting of $x$-$y$ planes of size 182 $\times$ 182~$\mathrm{\mu}$m and a resolution of 1024 × 1024 pixels was taken. The $z$ resolution was set to 3 micrographs per 1~$\mathrm{\mu}$m.

\subsection{Image processing and particle tracking}

The confocal micrographs were exported as two separate channels. Microparticle positions were located via a custom-written 3D detection algorithm in IDL (Exelis Vis, version 8.7). This algorithm detects particle edges and uses a Hough transform to find the centers, as reported in Bindgen~et al.~\cite{Bindgen2020}. To visualize the trajectory of individual particles during compression, 3D coordinates, detected at each compression step, were tracked based on a modified version of the IDL code by E.R.Weeks and J.C.Crocker~\cite{Weeks2000}. 

The network structure was analyzed by calculating the coordination number and clustering coefficient based on the positions of individual particles. The coordination number is the count of neighboring microparticles surrounding each particle, with contacts identified when the gap between their surfaces is less than 6 pixels or 1.08~$\mathrm{\mu}$m, taking into account the $z$-resolution errors and the measured particle radius. The clustering coefficient is defined as $c =\frac{2e}{z(z-1)}$, where $e$ is the number of connections between the neighboring particles and the $z$ is the coordination number~\cite{Bindgen2020}. The clustering coefficient has a value between 0 and 1, denoting a string-like or a fully clustered network, respectively. The clustering coefficient was determined using Python package NetworkX~\cite{Bindgen2020}.

The liquid channel stacks were first binarized using the default algorithm provided by ImageJ, after which the noise is removed using the despeckle command. The binary stacks were then analyzed using the combination of the implemented functions bwconncomp and regionprops3 provided by Matlab, where the volume and surface area of individual bridges were extracted. 

\section{Results and discussion}

\subsection{Effect of hydrophobicity on nanoparticle migration and distribution}

\begin{figure*}[tb]
\centering
  \includegraphics[width=0.8\textwidth]{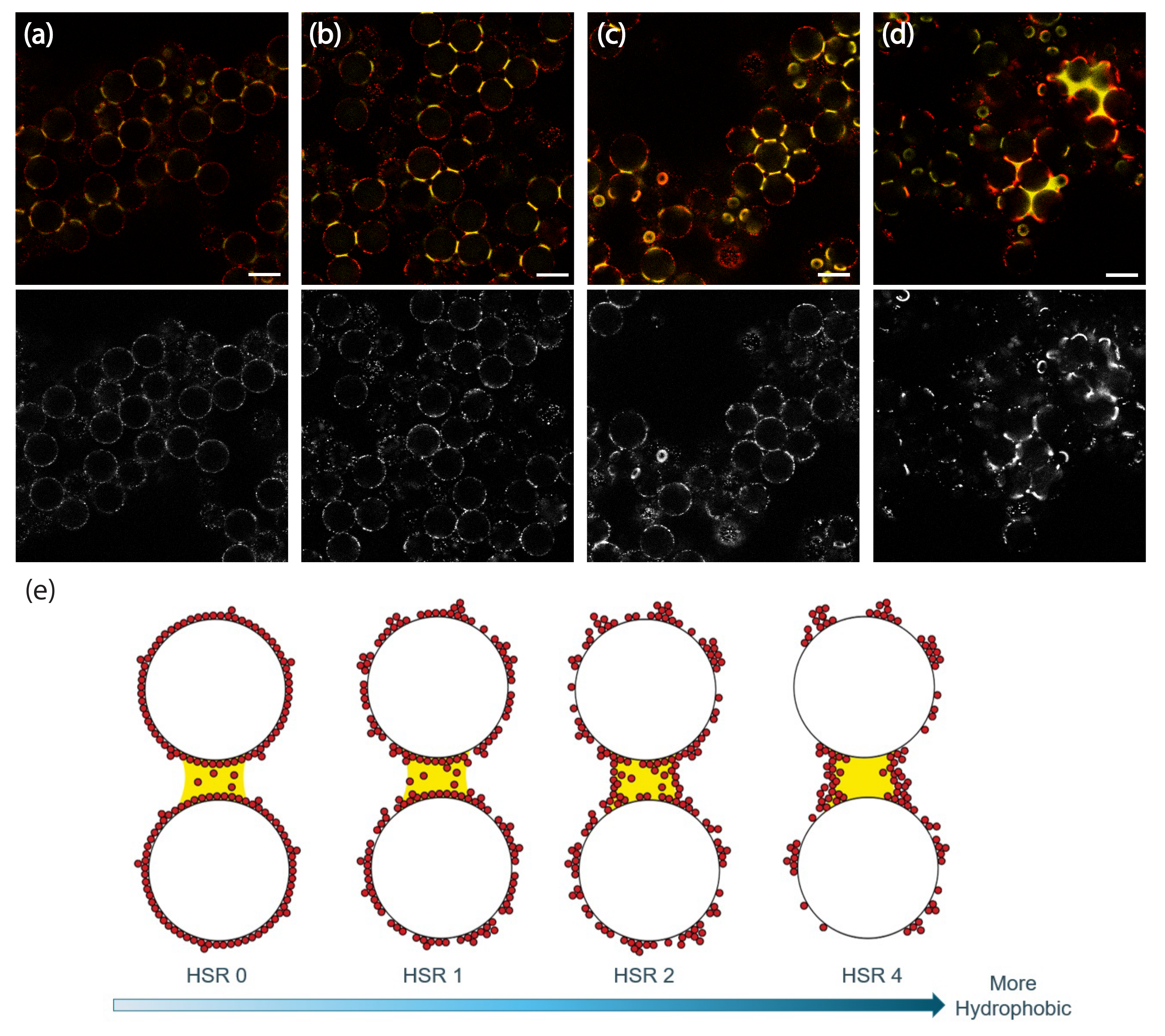}
  \caption [Confocal micrographs of capillary nanosuspensions with different nanoparticle hydrophobicities, schematic illustration of bridges and particles, and detected bridge sizes.]{(a)-(d) Confocal micrographs of capillary nanosuspensions ($\phi_{solid}$ = 20~vol\%, $\phi_{sec}$ = 1~vol\%, $\phi_{NP}=0.01$~vol\%). The nanoparticles, added in the secondary liquid phase, range from (a) untreated, hydrophilic particles, (b) HSR1, (c) HSR2, to (d) HSR4, the most hydrophobic particles.  The combined images are shown above with the secondary fluid in yellow and the nanoparticles in red. The nanoparticle channel alone is shown in the lower images. The scale bars are 10~$\mathrm{\mu}$m. (e) Illustration of migration of nanoparticle with different hydrophobicities in capillary nanosuspensions.}
  \label{fig: NPposition}
\end{figure*}

The present system consists of hydrophilic, spherical silica microparticles surface characterized silica nanoparticles (red) in an undyed bulk oil with aqueous glycerol (yellow) as the bridging liquid. The nanoparticle positions, relative to the microparticle and secondary bridges,  is shown in the confocal image slices of Fig.~\ref{fig: NPposition}a to \ref{fig: NPposition}d. 
The hydrophobicity of nanoparticles ($\phi_{NP}$=0.01~vol\%) increases from left (HSR0) to right (HSR4) with the nanoparticle channel shown alone in grayscale. In the wet state, the location of the nanoparticles is influenced by the nanoparticle hydrophobicity unlike in sintered materials where nanoparticles are found to be located solely at the contact region between microparticles~\cite{Weis2020, Park2019}.

HSR0 nanoparticles, i.e. NPs untreated with HMDS, deposit evenly on the microparticle surfaces (Fig.~\ref{fig: NPposition}a). The nanoparticle channel shows no difference in nanoparticle density near the capillary bridges. The hydrophilic nanoparticles do not preferentially stay inside the liquid bridges. The size and surface area of the secondary liquid images are as demonstrated in Supplemental Fig.~S1. A range of bridge sizes and shapes are present, but the majority of the bridges in HSR0 span small particles, binary or trimer. In Fig.~\ref{fig: NPposition}b (HSR1), the intensity at the capillary bridge region increases and the nanoparticle distribution on microparticle surfaces becomes more patchy. The HSR1 (blue) bridges have shifted to larger volumes (Fig.~S1) with a smaller total distribution area, where each population of bridges types~\cite{Scheel2008} are more separated. 
The counterintuitive mechanism of hydrophobically treated nanoparticles having a greater propensity to be trapped in the aqueous phase is intensified by HSR2 nanoparticles, as can be seen in Fig.~\ref{fig: NPposition}c. With the HSR2 treatment, the nanoparticle clusters are larger than HSR1, as evidenced by the higher intensity in the regions where NPs are present and the distribution of nanoparticles becomes more inhomogeneous, partially removing the circular outlines around the microparticles. Meanwhile, the nanoparticles are unevenly distributed in the bridges, which take on a bright toroidal structure. The bridges become wider for HSR2 and the distribution shifts towards bridges spanning larger numbers of particles. At HSR4, Fig.~\ref{fig: NPposition}d, the nanoparticles appear to present more neutrally wetting behavior, as the nanoparticle distribution becomes more uneven with minimum clusters on microparticle surfaces. Comparing the toroidal structures from HSR2 to HSR4, the rings found in the HSR4 sample are thinner and incomplete. The particles are expelled from the secondary liquid and the bridge size distribution overlaps with sample HSR0. The average nanoparticle cluster coverage rates on microparticles are 65.8~\%, 61.3~\%, 57.7~\%, and 50.7~\% for HSR0, HSR1, HSR2 and HSR4 samples respectively, in descending order.  

The migration mechanism of nanoparticles is summarized in Fig.~\ref{fig: NPposition}e. As hydrophobicity increases (left to right), the nanoparticles, adsorbed on microparticle surfaces, begin to become more aggregated and relocate towards liquid bridges, eventually ending up at the three-phase contact lines. For hydrophilic nanoparticles, the nanoparticles are well spaced between each other and attracted to the microparticles, likely due to the existence of hydration layers of up to 25~nm induced by dense surface silanol groups~\cite{Hamamoto2015, Sulpizi2012, Lu2014}. As the silanol group density decreases, the hydrophobized nanoparticles start to become surface active and repelled by the liquid films and the hydrophilic microparticle surfaces~\cite{Lu2014, Hamaker1937}. Meanwhile, nanoparticle hydration layers start to shrink and they begin to aggregate~\cite{Alvo2010}. The underlying rationale for this phenomenon is the transition from a van der Waals driven attraction of the nanoparticles to the microparticles, to the situation where nanoparticle-nanoparticle attraction and interfacial adsorption dominate. During this transition, it has been observed that the nanoparticles migrate towards the liquid bridges (HSR0 to HSR4). When the nanoparticles are moderately hydrophobized (HSR1 and HSR2), a fraction of them tend to stay at three-phase contact lines~\cite{Lu2014, Bresme2007}. Considering the nanoparticles are pre-wetted by the secondary liquid, the aggregates deposited at the contact lines expand the diameters of the liquid bridges, increasing average bridge volumes. Hydrophobic HSR4 nanoparticles however, do not contribute to the growth in bridge volume since they are primarily expelled from the secondary liquid, bringing the distribution back to the starting position (HSR0). This effect is explained by the increase in lateral capillary immersion force, $F_{\text {im}}$, which is attractive when the wettability of both nanoparticles are the same, resulted from the same menisci behavior~\cite{Bresme2007, Kralchevsky1994}. From the fluid dynamics point of view, as nanoparticles become more hydrophobic, the surface energy decreases, resulting in an increase in the hydrodynamic slip length~\cite{Ortiz2013}. Due to slip between liquid molecules and nanoparticles, an increase in the mismatch in wettability makes the nanoparticles less likely to move together with secondary liquid during mixing, resulting in patchy agglomerated nanoparticle clusters~\cite{Anyfantakis2017}.

\subsection{Rheological measurements and gap dependence}\label{sec: rheo}

\begin{figure}[tb]
\centering
  \includegraphics[width=0.5\textwidth]{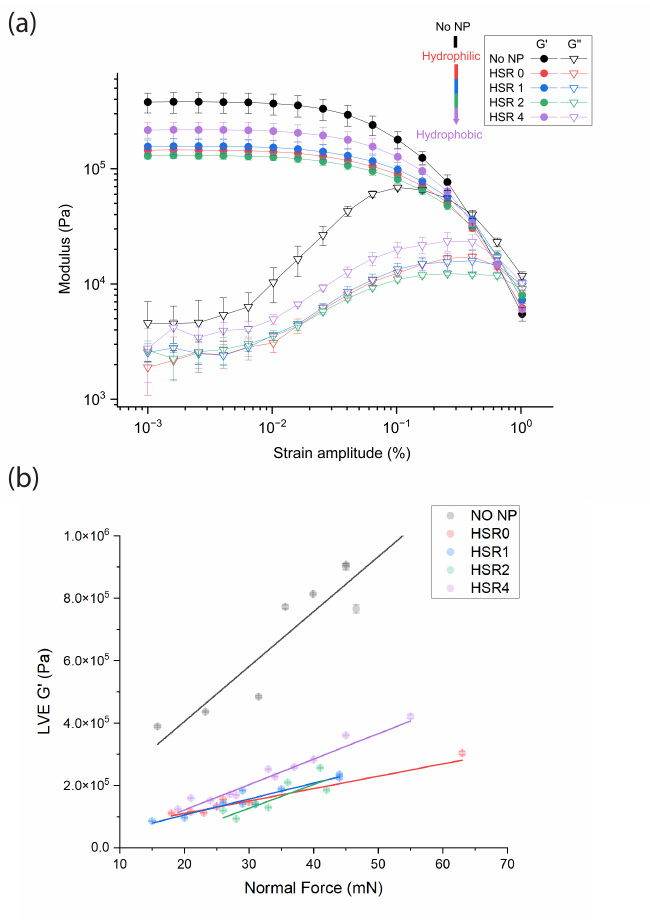}
  \caption{(a) Rheological measurements of capillary suspensions without nanoparticles and capillary nanosuspensions with nanoparticles of different hydrophobicities (b) The storage modulus in the LVE region ($G'_0$) versus individual normal force at the trimming position (1.025 mm). The LVE moduli are given by the corresponding average using a 5~\% criteria. The error bars are sometimes smaller than the symbols. Linear fits are shown to illustrate trends.}
  \label{fig: hydrophobrheo}
\end{figure}

Without freely moving nanoparticles, capillary suspensions are generally stable and rigid with the high shear moduli~\cite{Koos2011, Bindgen2020, Bossler2018}. Differently hydrophobized nanoparticles have a surprisingly large impact on their rheological measruements. This is shown in the oscillatory amplitude sweeps shown in Fig.~\ref{fig: hydrophobrheo}a.

The samples without nanoparticles have the highest storage modulus in the linear viscoelastic (LVE) region ($G'_0 = 3.8 \pm 0.7 \times  10^5$~Pa) with a relatively steep increase in the loss modulus $G''$ with strain amplitude, showing a peak at $\gamma$ = 0.1 \%. The crossover point ($G'=G''$) is at $\gamma$ = 0.4 \%. The storage modulus decreases by half an order of magnitude with only 0.01~vol\%  nanoparticles added with the secondary liquid. The samples with nanoparticles (colored) generally behave similarly, with slightly delayed $G''$ peaks at $\gamma$ = 0.4 \% and crossover nearby $\gamma$ = 0.6 -- 1 \%, despite the striking morphological differences. HSR4 shows an intermediate behavior with a higher moduli than the other nanoparticle-laden samples. This corresponds to the phenomenon shown in Fig.~\ref{fig: NPposition}d, where HSR4 nanoparticles are located at the interface of the bulk and secondary liquids. Surprisingly, the rheological response shows little dependence on the bridge size (Fig.~S1 where HSR0 and HSR4 have similar liquid bridge size distribution and HSR2 has the lowest shear moduli despite having the highest average bridge size. This is unexpected since the shear moduli should correspond to the influence of the bridge size on the capillary force in the limit of small bridges~\cite{Bossler2016, Bindgen2020, Bossler2017, Koos2011}.

The loss modulus $G''$ represents energy dissipation during network rearrangement~\cite{Bossler2018, Donley2020, Bindgen2020}, and is lowered due to the presence of nanoparticles with the minimally retarded crossover point. The decrease of the loss modulus is possibly related to the reduction of Hertzian contact between microparticles~\cite{Natalia2022} and the shift in $\gamma$ can be caused by network flexibility induced by nanoparticles where bridges are prone to break at a slightly higher strain~\cite{Donley2019}. The higher loss modulus for the HSR4 sample would also be consistent with this change in the Hertzian contact since the fewer particles in the bridge would interfere less with direct microparticle-microparticle contacts.

During rheological experiments, it was discovered that the measured shear moduli of capillary (nano)suspensions fluctuated with the loaded sample volume even if the excess is removed prior to the measurement. The normal force increases rapidly as the top plate is lowered to the desired gap, as shown in Fig.~S2. Upon compression, the normal forces increases gradually until  $F_\text{N} \approx$ 10 mN, after which the dependence on the gap becomes more pronounced. This dependence on the gap differs for the samples with and without nanoparticles. Despite having the lowest $F_\text{N}$ at beginning of compression, the normal force for the sample without nanoparticles rises rapidly as the plate approaches the trim gap. A similar effect has been reported by Allard~et~al., possibly owing to the increase in effective volume of microparticles during compression as bulk liquid is expelled from the gel network~\cite{Allard2022}. 

The storage modulus in LVE region as a function of the normal force at the trimming position (1.025~mm, 25~$\mathrm{\mu}$m above the measurement gap) is plotted in Fig.~\ref{fig: hydrophobrheo}b. For all the samples, the fitted correlation between $G'_0$ and normal force is linear. The slope $G'_0/F_N$ reflects the rigidity of the network upon being compressed. The rigidity is the highest for the sample without nanoparticles with a slope of $G'_0/F_N = 1.8\pm 0.1 \times 10^4$~Pa/mN and decreases for the sample with nanoparticles. For HSR0, the slope of $4.0 \pm 0.9 \times 10^3$~Pa/mN is the lowest. With increasing hydrophobicity, the slope rises incrementally up to $8.2 \pm 1.0 \times 10^3$~Pa/mN for HSR4. However, the HSR4 samples can only be differentiated in $G'_0$ from HSR0, HSR1 and HSR2 at regions with higher normal forces ($F_N> 30$~mN).

\subsection{Microparticle movement during compression}

To study microstructural changes of capillary (nano)suspensions during the lowering of the rheometer geometry, a step-by-step compression test was conducted on a confocal microscope. Due to the intrinsic rigidity of capillary suspension~\cite{Koos2011, Bossler2017, Bossler2018, Bindgen2020, Allard2022}, the gel structure is not microscopically smooth after being sheared by a doctor blade. Hence, the highest point of the network was first centered in the x-y plane of the 3D confocal images. 

\begin{figure*}[tb]
\centering
  \includegraphics[width=0.8\textwidth]{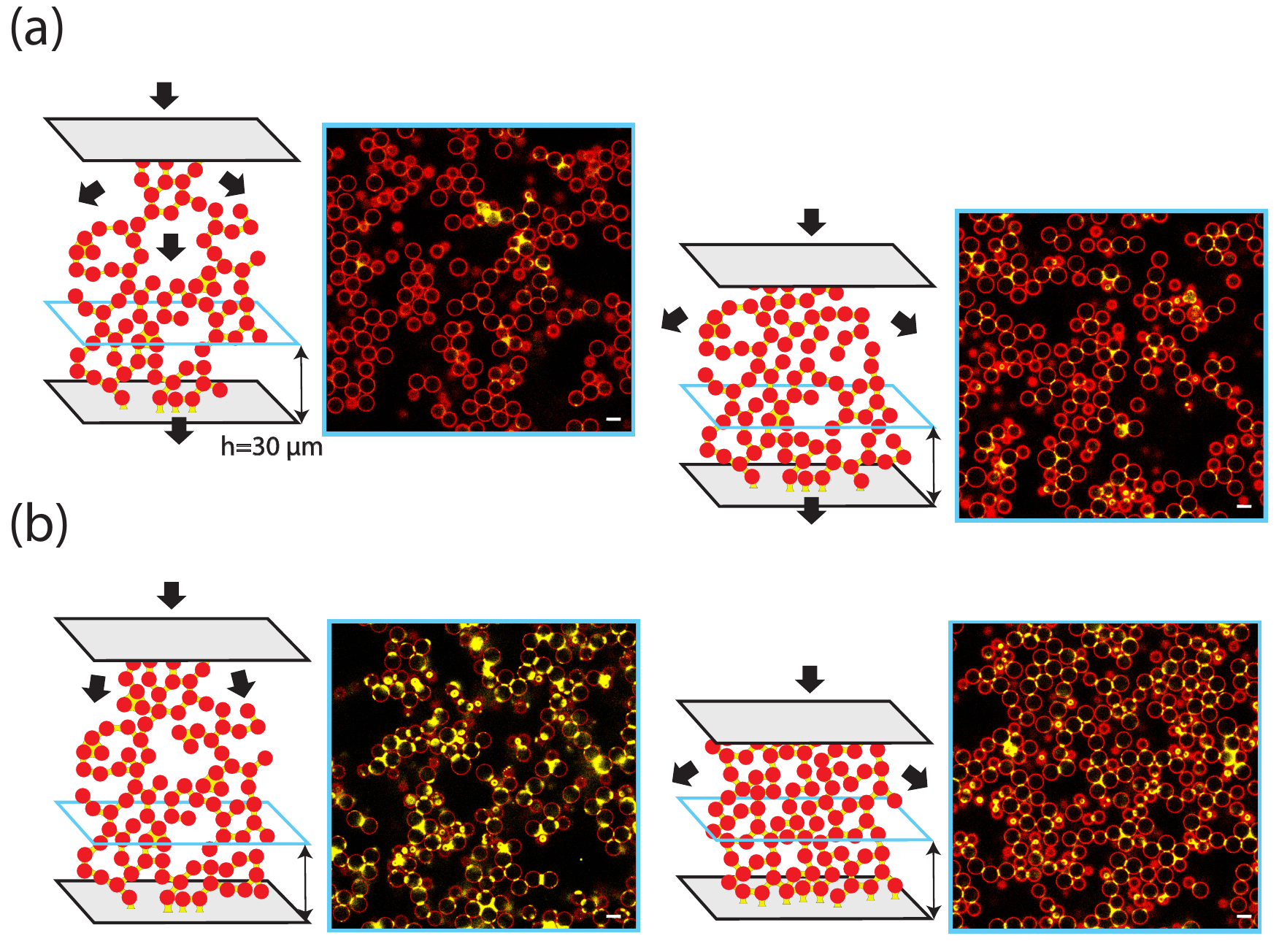}
  \caption{Illustration and confocal micrographs (left) before and (right) after compression for capillary suspensions (a) without nanoparticles, and (b) with HSR0 nanoparticles. Gray rectangle planes represent the top and bottom geometry of the shell cell (glass slides), microparticles are displayed in red and secondary fluid bridges are in yellow. The black arrows represent the approximate direction of the force chain. The cyan frames indicate the planes where micrographs were taken, with a consistent height of 30~$\mathrm{\mu}$m from the current bottom layer. The scale bars are 10~$\mathrm{\mu}$m.}
  \label{fig:CompressionAnime}
\end{figure*}

The results of the compression test are shown in Fig.~\ref{fig:CompressionAnime}. 
For the sample without nanoparticles (Fig.~\ref{fig:CompressionAnime}a), only a few larger, coalesced secondary liquid bridges (yellow) are visible before compression. After compression, single bridges are visible using the same microscope settings, where the bridge number increases and average size decreases ($V$ = 41.8 $\mathrm{\mu}$m$^3$ before and 36.1~$\mathrm{\mu}$m$^3$ after compression). The corresponding beginning of the highest quartile remains stable with a number of 34.51~$\mathrm{\mu}$m$^3$ and 36.17~$\mathrm{\mu}$m$^3$.The microparticle fraction remained roughly the same, as shown in the separate channels displayed in Fig.~S3, with a particle coverage of 39.2 \% and 44.1 \% for before and after compression, respectively. During the displacement of the top plate, the bottom slides flex, displacing the observed bottom position by 21~$\mathrm{\mu}$m for samples without nanoparticles (Fig.~\ref{fig:CompressionAnime}a) whereas nearly no deflection (< 2~$\mathrm{\mu}$m) was observed when 0.01~vol\% HSR0 nanoparticles were well distributed on microparticle surfaces, as shown in Fig.~\ref{fig:CompressionAnime}b. After full compression, the microparticle volume fraction $\phi_\text{MP}$ increases significantly with secondary liquid redistribution, leading to smaller average bridge size, and the whole distribution shifts towards the left on x-axis. The average bridge volumes $V$ = 85.8~$\mathrm{\mu}$m$^3$ before and 43.9$\mathrm{\mu}$m$^3$ after compression. The beginning of the highest quartile decreases from 64.93~$\mathrm{\mu}$m$^3$ to 44.07~$\mathrm{\mu}$m$^3$.

\subsubsection{Network response to compression}

As the top glass of the shear cell is lowered, the sample height decreases, but the change in the gap is not directly equal to the applied deformation of the top plate, due to the deformation of the bottom plate. 
\begin{figure}[tbp]
\centering
  \includegraphics[width=0.5\textwidth]{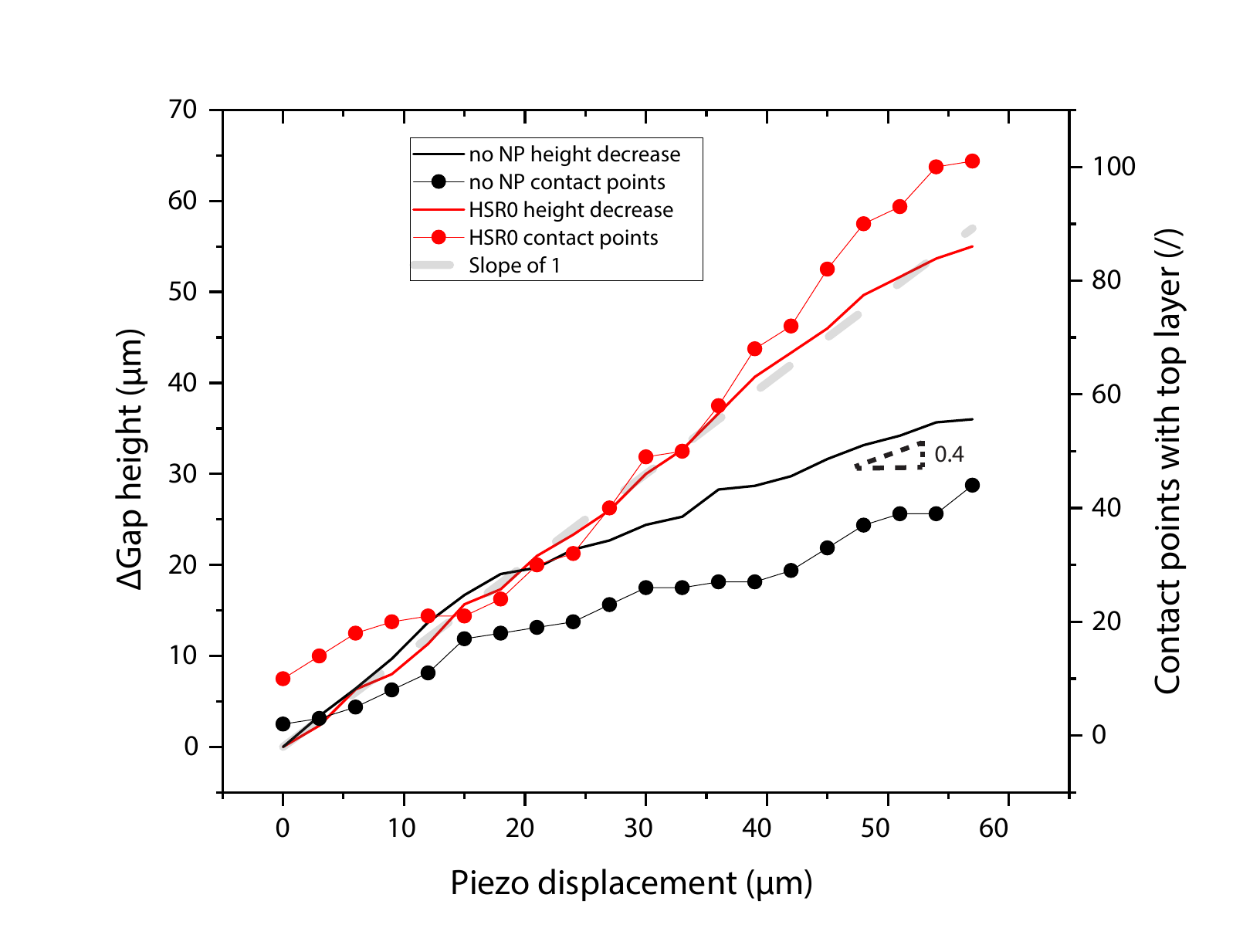}
  \caption{Relative sample compression and corresponding changes in the number of contact points between microparticles and top layer cover glass versus applied deformation of the top plate.}
  \label{fig:piezolowering}
\end{figure}
As shown in Fig.~\ref{fig:piezolowering}, a shift in both the rate of sample compression and the number of microparticle contact points between the sample and top plate can be divided into two regions.
In the first 21~$\mathrm{\mu}$m of compression (Stage I), the network rearrangement caused by compressing a limited amount of microparticles whereas more collective behavior is captured in Stage II.

\paragraph*{Stage I}

In Stage I, the decrease in the gap follows the applied compression (slope of 1), as shown in Figure~\ref{fig:piezolowering}. Due to the height variation of the local microparticle network, the number of initial contact points between the top glass and microparticles is limited. Although the number of contact points increases almost linearly with piezo displacement for both samples, the increasing speeds in contact points are different at the beginning stage, likely due to the topology of the initially uneven upper surface. 

\begin{figure*}[tb]
\centering
  \includegraphics[width=0.95\textwidth]{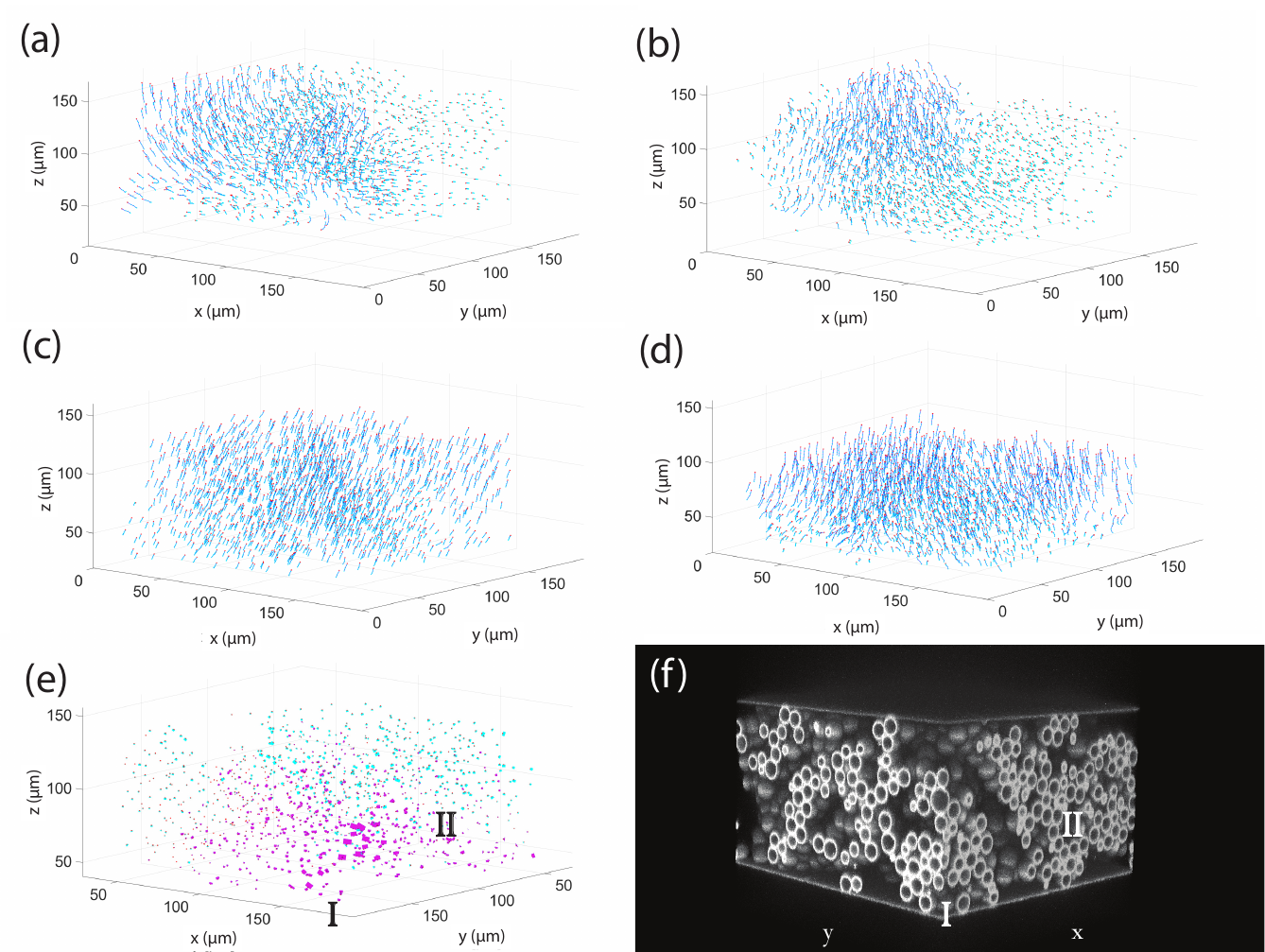}
  \caption{Microparticle movement during compression during Stage I for the sample (a) without nanoparticles, (b) with HSR0 nanoparticles and during Stage II for the sample (c) without nanoparticles, and (d) with HSR0 nanoparticles. Red dots represent the starting point of microparticles and particle positions at each step are denoted by cyan dots, internally connected by blue lines. The sample without nanoparticles showing (e) relative movement of the microparticles and (f) the initial 3d image. The cyan trajectories in \textit{e} denote particles exhibiting more downward motion than the average and magenta trajectories represent the microparticles that move in a less than the average. The line widths are proportional to the distance and red dots are the starting point of microparticles.} 
  \label{Fig: 3d}
\end{figure*}

To capture the difference in the network response, the particle positions are tracked throughout the compression and their displacement is plotted in Fig.~\ref{Fig: 3d}. 
For both the sample without nanoparticles (Fig.~\ref{Fig: 3d}a) and with HSR0 NPs (Fig.~\ref{Fig: 3d}b), the displacement vectors of each microparticle  are strongly influenced by the neighboring particles in Stage I. The particles at the top layer are displaced downwards and this displacement is propagated through both networks towards the bottom plate. The motion is highly non-uniform with large numbers of microparticles in both samples showing little to no displacement in Stage I. The vertical motion is both damped and redirecting near the bottom layer. Lateral displacements are observed for both samples indicating flow of the microparticles away from the predominant region of compression. The lateral displacement is more prevalent in the sample without NPs (Fig.~\ref{Fig: 3d}a) where the particles located near (0, 0) move in a positive x-direction and the particles in the lower area of the graph (150~$\mathrm{\mu}$m, 0) where several particles are displaced upwards. The movements of the microparticles in the top layer are predominantly vertical and are damped as the force is propagated downward in the sample with NPs (Fig.~\ref{Fig: 3d}b). This particle motion is also reflected in the change in the microparticle volume fraction $\phi_\text{MP}$ (Fig.~S4). In the first seven steps, $\phi_\text{MP}$ increases in the upper layers for both samples, whereas it shows less change in the lower levels. This is particularly true for the sample with NPs where $\phi_\text{MP}$ of the lower layers remains constant until approximately 10 compression steps (30~$\mathrm{\mu}$m).

\paragraph*{Stage II}
After a displacement of the top piezo plate of approximately 21~$\mathrm{\mu}$m, the uneven initial structure is smoothed and an uniform force is exerted on the top layer of each sample. Thus, Stage II shows more homogeneous behavior; for both samples (Fig.~\ref{Fig: 3d}c and \ref{Fig: 3d}d), nearly every particle is displaced downward. In the sample with NPs (Fig.~\ref{Fig: 3d}d), the change in the gap height closely follows the displacement of the top plate (gray dashed line in Fig.~\ref{fig:piezolowering}) whereas the change in the gap height is reduced to a slope of 0.4 for the sample without NPs. This is also reflected in the change in the volume fraction (Fig.~S4 where the increase in $\phi_\text{MP}$, i.e. the actual compression of the sample, is tempered for the sample without nanoparticles. The bottom part of the shear cell is composed of a cover glass of thickness 130~$\mathrm{\mu}m$, which can flex under external forces~\cite{Timoshenko1959}. The maximum local deflection of centrally loaded plates with simply supported edges is given by Eq.~\ref{eq:plate},
\begin{equation}\label{eq:plate}
  y_{\max }=\frac{-\alpha W b^2}{E t^3}
  \end{equation}
where the load is $W$, $b$ is the length of the shorter edge, $\alpha$ is the shape ratio coefficient, $E$ is Young's modulus and $t$ is the thickness of the cover glass~\cite{Young2012}. In this set up, a local deflection of about 21~$\mathrm{\mu}m$ requires a load of 30 mN. As shown in Fig.~S2, this force is on the same order of magnitude of the normal force measured for decreasing gaps on the rigid rheometer and the normal force used for the oscillatory measurements (Fig.~\ref{fig: hydrophobrheo}a). This indicates that the transition from heterogeneous to homogeneous compression is related to the formation of force chains spanning the gap.

To eliminate the effect of the bottom plate deflection, the average particle movement has been subtracted from the particle traces in Fig.~\ref{Fig: 3d}e, noting that the view in Fig.~\ref{Fig: 3d}e has been rotated from that in Fig.~\ref{Fig: 3d}c. Particles with cyan trajectories move downwards more than the average whereas magenta denotes particles that moved downward less than the average (upwards relative to the average motion). Thus, the particles with a magenta trajectory resisted the applied compression more than those with a cyan trajectory. The line thickness in Fig.~\ref{Fig: 3d}e is scaled by the magnitude of the displacements. The same view of the initial 3d structure of the sample without NPs is  shown in Fig.~\ref{Fig: 3d}f. The larger number of magenta trajectories in the corner (150~$\mathrm{\mu}$m, 150~$\mathrm{\mu}$m) corresponds to the compact structure (cluster I). Such compact structures exhibit rigid behavior~\cite{Bindgen2020, Allard2022}. Meanwhile, cluster II (100~$\mathrm{\mu}$m < x < 150~$\mathrm{\mu}$m) is located above the large void and shows a more uniform distribution of cyan trajectories. This compact cluster moves consistently to fill the void. Therefore, the structural rearrangement of sample without nanoparticles is limited by the solid-like movement of the dense clusters~\cite{Bindgen2020, Bossler2018, Allard2022}.

For sample with HSR0 nanoparticles, almost no deflection at the bottom slide is observed. This implies that the particle rearrangement occurs to a much larger degree, as shown in the rapid increase in the microparticle volume fraction in Fig.~S4.  From the seventh compression step (21~$\mathrm{\mu}$m), the increase rate in the contact points with the top plate displacement accelerate (Fig.~\ref{fig:piezolowering}) and the streamlines are not unidirectional (Fig.~\ref{Fig: 3d}d). The particles at the top layers spread isotropically downwards, pushing other particles away (Fig.~S5). Thus, despite the higher number of contact points between the microparticles and the top plate, the force propagation between the slides is damped or `lubricated' for the sample with nanoparticles. In subsequent experiments, it is observed that the sample with NPs can be compressed to a microparticle volume fraction $\phi_\text{MP} \approx$ 40 vol\% after which the bottom glass deflects.

\subsubsection{Structural variation}

\begin{figure}[tbp]
\centering
    \includegraphics[width=0.5\textwidth]{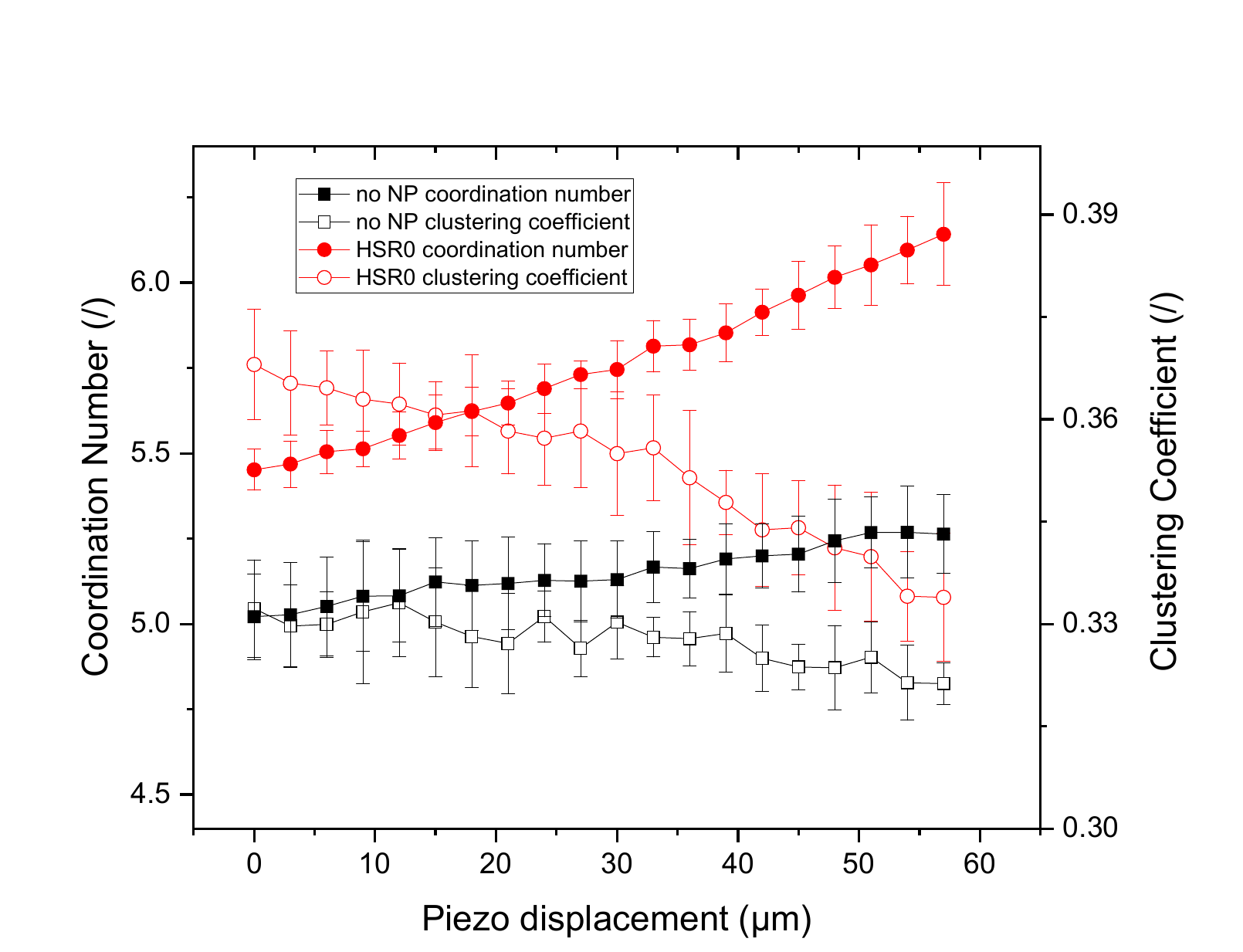}
  \caption{Average coordination number and clustering coefficient changes during compression. }
  \label{Fig: ActualAvgLayer}
\end{figure}

Before compression, the initial microparticle volume fraction for the samples both with and without nanoparticles are distributed around 24.31 $\pm$ 0.043 \% and 23.45 $\pm$ 0.025 \%, respectively, as shown in Fig.~S4. For the sample without nanoparticles (Fig.~S4a), most of layers exhibit minor increases upon compression with an average slope of 0.002 \% per step. With the addition of HSR0 NPs, the movement of microparticles increases with a slope of 0.008 \% per step (Fig.~S4b), converging to $\phi_\text{MP} \approx 30$~vol\%. The increase in relative movement is also demonstrated in Fig.~\ref{Fig: ActualAvgLayer}
where the average coordination number for that sample without nanoparticles increase slightly from 5.02 $\pm$ 0.13 to 5.26 $\pm$ 0.11, owing to the bulky response of microparticle cluster network~\cite{Bindgen2022}. When nanoparticles are added, the sparse clusters become compact upon compression, increasing the coordination number from 5.45 $\pm$ 0.06 to 6.14 $\pm$ 0.15. The distribution in the increase in average coordination number is as plotted in Fig.~S6, where the number of particles with high coordination numbers (dark red) increased more when nanoparticles are present (Fig.~S6b and S6d) whereas the sample without nanoparticles showed little change (Fig.~S6a and S6c). Two micrograph slices with superimposed location detection are shown in Fig.~S6e and S6f. The circular slice in each frame, here separated by a $z$-depth of 0.33~$\mathrm{\mu}$m, is shown in cyan and sphere's center indicated in the $z$-axis frame closest to its detected location. The location of the neighboring particles can be applied to calculate the coordination numbers and clustering coefficients, as described in Bindgen et al.~\cite{Bindgen2020}. 

The distribution of both the particle coordination number and clustering coefficient is also more uniform for the sample with nanoparticles (Fig.~S7).  Unhindered cluster densification and restructuring implies that the presence of nanoparticles results in a lower rotational energy scale of microparticles~\cite{Brodu2015, Nguyen2019}. That is, the nanoparticles serve to lubricate the contacts. The change in the flexibility of clusters is also evident in the distribution of bridge sizes and shapes, as shown in Fig.~S8. After compression, the sample with nanoparticles exhibits more dimers and few trimers than before compression for both the sample with and without nanoparticles as shown by the reduction in the points in the upper right quadrant (large bridge sizes with high surface area). 

\begin{figure}[tbp]
\centering
  \includegraphics[width=0.4\textwidth]{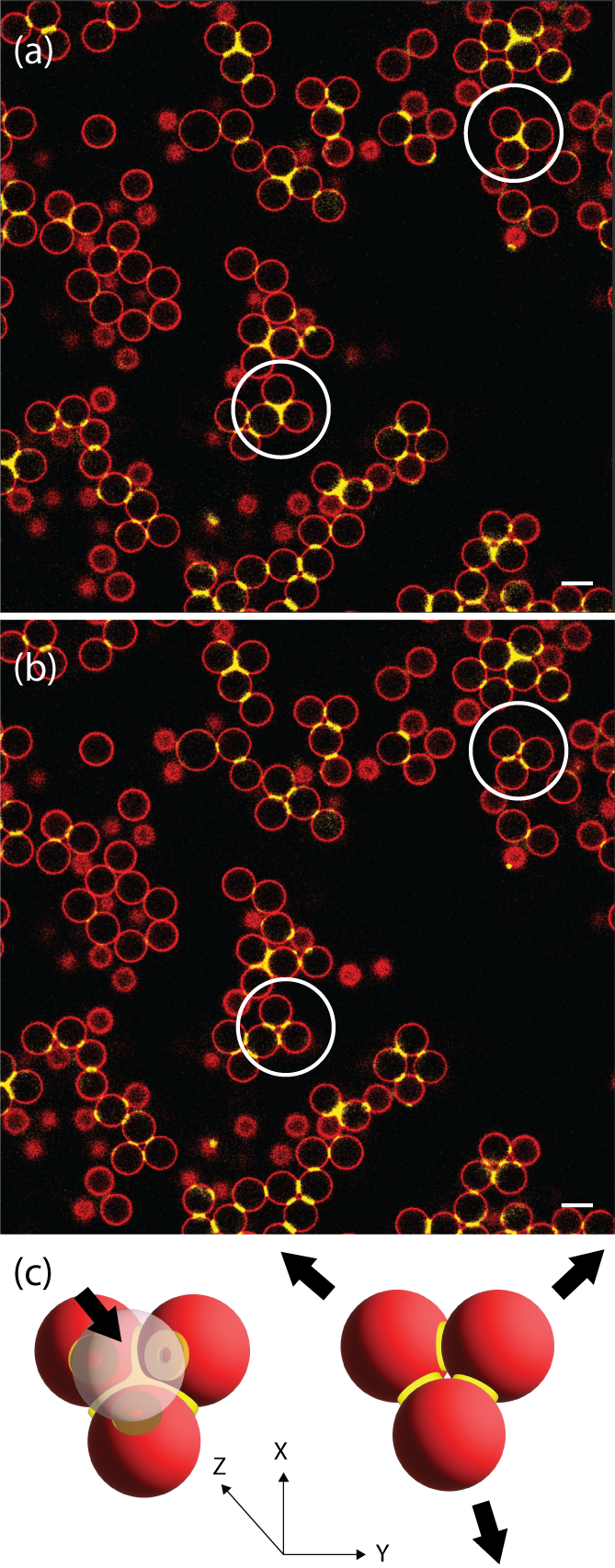}
  \caption{(a-b) Consecutive confocal micrographs (x-y plane) during compression of a sample without nanoparticles showing trimer breakage, as illustrated by the sketch (c). The scale bars are 10~$\mathrm{\mu}$m}
  \label{fig:trimerbreakage}
\end{figure}

For the sample with nanoparticles, the initial coordination number is higher than the sample without nanoparticles, possibly caused by the unavoidable minor compression due to the pre-shear prior to the experiment. During compression, both the coordination number and clustering coefficient fluctuate around their initial values, with a slight tendency to increase and decrease, respectively (Fig.~\ref{Fig: ActualAvgLayer}). Elevated coordination numbers are usually associated with the merging of clusters with secondary liquid bridges binding all of the particles~\cite{Bindgen2020, Bindgen2022, Allard2022}, but the reduction in the clustering coefficient implies that another mechanism may be at work. As can be observed in the images from consecutive steps before (Fig.~\ref{fig:trimerbreakage}a) and after (Fig.~\ref{fig:trimerbreakage}b) a 3~$\mathrm{\mu}$m compression.  

The lower three particles of a tetradedron (circled) become separated into three dimers. This process is illustrated in the sketch (Fig.~\ref{fig:trimerbreakage}c), where the upper particle pushes the lower three apart to accommodate the lowering of the upper particle.
Meanwhile, elevated distances between microparticle surfaces break the direct contact between microparticles, leading to the slight reduction in the clustering coefficient, as shown in Fig.~\ref{Fig: ActualAvgLayer}. This breakage of a funicular trimer into pendular bridged dimers increase the total capillary force~\cite{Wang2017} and results in further resistance of the capillary suspension network without nanoparticles against compression. 

For sample with hydrophilic HSR0 nanoparticles, the bridge size prior to compression is much higher, with majority of the population lying between 2 - 40~$\mathrm{\mu}$m$^3$ (52~\%), possibly due to the homogenization process being hindered by the presence of nanoparticles (Fig.~S8b). During compression, the number of funicular clusters is reduced such that the sample after compression shows a majority of binary bridges (Fig.~\ref{fig:CompressionAnime}b). The bridge size distribution is narrowed to 66 \% in the range of 2~$\mathrm{\mu}$m$^3$ < $V_\text{cb}$ < 40~$\mathrm{\mu}$m$^3$. The deagglomeration of funicular clusters, induced by nanoparticle films enhancing liquid redistribution~\cite{Bossler2016, Wang2017}, is associated with a decrease in the clustering coefficient from 0.368 $\pm$ 0.008 to 0.334 $\pm$ 0.009 in the present case.

\subsection{Role of nanoparticles} \label{sec: NProle}

During compression, the nanoparticles elevate relative movement between microparticles and induce a more homogeneous network with a narrower bridge size distribution. The changes are indicate the promotion of liquid exchange between bridges and the reduction of Hertzian contact force.

\subsubsection{Promotion of secondary liquid exchange}

\begin{figure}[tbp]
\centering
  \includegraphics[width=0.4\textwidth]{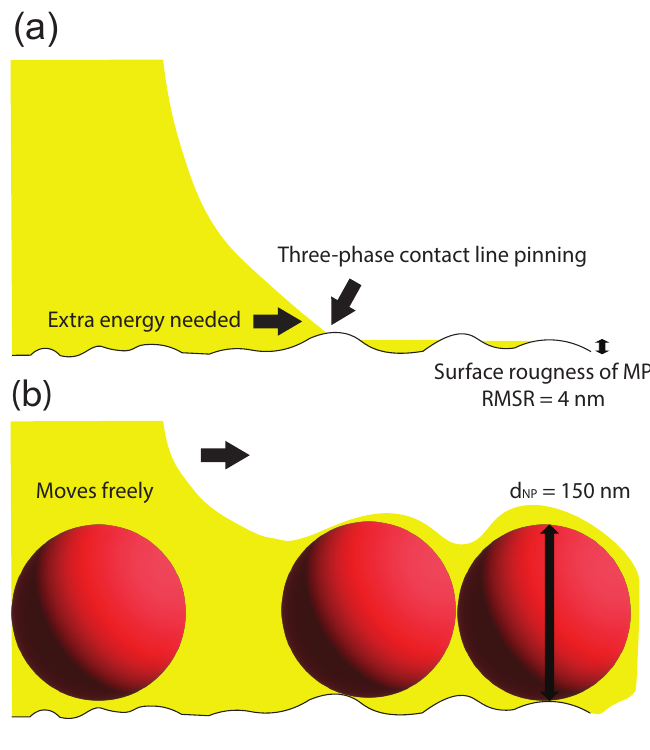}
  \caption[Illustration of contact line pinning effect of liquid bridges at three-phase contact lines.]{Illustration of contact line pinning effect of liquid bridges at three-phase contact lines (a) without and (b) with nanoparticles. The core-shelled nanoparticles here are with an average diameter of 150~nm, with 90~nm fluoresently labeled core and an undyed shell of thickness roughly 30~nm. }
  \label{fig:contactlinepinning}
\end{figure}

The number of visible bridges increases significantly when nanoparticles are incorporated in the secondary liquid (Fig.~S3). By changing the hydrophobicity of the nanoparticles, the liquid bridge distribution is modified. As shown for the hydrophilic HSR0 nanoparticles, these large bridges tend to be redistributed into smaller bridges as evidenced by the narrower bridge size distribution (Fig.~S8). The capillary bridges for the sample without NPs do not experience this phenomenon. One explanation for this difference, shown in Fig.~\ref{fig:contactlinepinning}a, 
is the elimination of contact line pinning~\cite{Bossler2016, Wang2017} due to coverage of the nanoparticles by the secondary liquid. As shown in Fig.~\ref{fig: NPposition}a, the HSR0 nanoparticles are evenly distributed around microparticle surfaces without apparent influence of the NPs and on the secondary liquid (or vice versa). 

When micrometer-sized bridges are sheared on pinned surfaces, the lateral adhesion force must overcome contact angle hysteresis to initiate sliding~\cite{Gao2018, Song2021}. The lateral adhesion force $F_\text{lateral}$ is given by
\begin{equation}\label{eq:Fad}
  F_\text{lateral}=\gamma L_\text{s} \left(\cos\theta_\text{s}^\text{Rear} - \cos\theta_\text{s}^\text{Front} \right)
  \end{equation}
where $\gamma$ is the liquid-liquid interfacial tension, $L_\text{s}$ is the contact width, and $cos\theta_\text{s}^\text{Rear}$ and $cos\theta_\text{s}^\text{Front}$ are the front and rear contact angles in static (pre-sliding) regime, respectively~\cite{Gao2018}. The rear and front contact angles are generally a material property and difficult to determine on a micrometer-sized scale due to the limited pixel resolution of the confocal micrographs~\cite{Gao2018, Song2021, Pilat2012}.

Based on the sliding model of Gao et al.~\cite{Gao2018} for water on a silicon wafer, adjusted by the interfacial tension, $F_\text{lateral}$/$L_\text{s} \approx$ 20~nN/$\mathrm{\mu}$m. The capillary bridge neck diameter (contact width $L_\text{s}$) for single capillary bridge contact is generally between 2~$\mathrm{\mu}$m and 4.5~$\mathrm{\mu}$m (3~$\mathrm{\mu}$m$^3$ < $V_\text{cb}$ < 20~$\mathrm{\mu}$m$^3$). Therefore, the $F_\text{lateral}$ is between 40 and 90 nN for capillary bridges between two 10~$\mathrm{\mu}$m microparticles. If it is assumed that the vertical force, resulting in the deflection of the glass, is evenly distributed to each particle contacting the upper layer, each particle experiences approximately 200 nN at the last compression step. In a tetrahedral arrangement, e.g. Fig.~\ref{fig:trimerbreakage}c, the applied force in the horizontal direction is 65 nN for the lower particles. This value falls within the limit threshold for sliding with the caveat that particle roughness would increase this value over that of silicon. This calculation aligns with the hypothesis that the lateral force ($F_\text{lateral}$) predominantly contributes to the rigidity of the network without nanoparticles.

By introducing freely moving nanoparticles, the observable thin liquid films surround the surface of the microparticles, as illustrated in Fig.~\ref{fig:contactlinepinning}b, are able to internally connect separate bridges, promoting liquid exchange and narrowing the bridge size distribution. Since the nanoparticle diameter is much larger than than the particle asperity of 4~nm~\cite{Allard2022}, the induced films are able to nearly eliminate the extra energy required to wet microparticles~\cite{Hntsberger1981}. Therefore, $F_\text{lateral}$ becomes negligible in the HSR0 system, eliminating the contact angle pinning and contact angle hysteresis. 

The decreased in $G'$ for the samples with nanoparticles (Fig.~\ref{fig: hydrophobrheo}a), can also be explained by the reduction in $F_\text{lateral}$. The increase in $G'$ for the HSR4 sample would also be consistent with its partial return when the nanoparticles adsorb to the liquid-liquid interface rather than predominantly being located at the microparticle surface (Fig.~\ref{fig: NPposition}d). Since the elasticity modulus $G'$ in the LVE region is predominantly related to the contact number between microparticles, which increases linearly during compression (Fig.~\ref{Fig: ActualAvgLayer}), the linear correlation between $G'$ and the normal force  (Fig.~\ref{fig: hydrophobrheo}b).

\subsubsection{Reduction of Hertzian contacts}\label{sec: hertzian}

In capillary suspensions, the capillary force ensuring particle contact deforms the microparticles on a nanometer scale, resulting in a Hertzian contact force~\cite{Natalia2022, Wang2017b}. When incorporating nanoparticles between microparticles inside the capillary bridges, the microparticle deformation is reduced. With increasing hydrophobicity from HSR0 to HSR4, the nanoparticles are displaced from the microparticle contact points (Fig.~\ref{fig: NPposition}), restoring the Hertzian deformation. This observation corresponds to the curves shown in Fig.~\ref{fig: hydrophobrheo}b, where the $G'/F_\text{normal}$ slope value increases slightly with increasing nanoparticle hydrophobicity until a maximum value is reached for sample without nanoparticles. 
The effect of friction force provided by Hertzian contact is more pronounced for the loss modulus $G''$ in the medium amplitude oscillatory shear (MAOS) regime~\cite{Natalia2022, Ewoldt2013}. As shown in Fig.~\ref{fig: hydrophobrheo}a,  the $G''$ of samples without nanoparticles shows a pronounced peak whereas the peaks in the NP samples are more subdued. 

A similar decrease in $G'$ in the linear viscoelastic region $G'_0$ by half an order of magnitude. This effect is more pronounced for hydrophilic nanoparticles than the hydrophobized nanoparticles that more preferentially adsorb to the liquid-liquid interface. The nanoparticles also reduce the peak in the loss modulus $G''$. A similar decrease in $G'$ was observed  when nanoparticles are fixed on the microparticles~\cite{Allard2022} due to the extra liquid required to fill the resulted asperities. In the rough particle system, however, there was no observed decrease in the $G''$ peak and a decrease in microparticle degree of freedom, triggered by the interlocking asperities, is reflected by the increase in critical strains. 

The combination of liquid exchange promotion and Hertzian contacts reduction reduces the influence of $G'_0$ on the normal force and elevated relative movement of the microparticles. This enhanced flexibility is shown using 3D particle tracking during compression experiments where the network incorporating nanoparticles is more readily compacted and restructured.

\section{Conclusions}

Via confocal microscopy, we visually determined the location of differently hydrophobized silica nanoparticles, added with the secondary fluid in capillary suspension precursor systems. When they nanoparticles are hydrophilic, they are  found to be preferentially deposited on microparticle surfaces. With increasing hydrophobicity, they aggregate, creating a patchy deposition on the microparticles, and are further adsorbed to the liquid-liquid interface. This observation is counter to previous work with porous ceramics where nanoparticles are solely found at the contact regions between microparticles after sintering~\cite{Weis2020,Park2019,Dittmann2016}. Thus, either the particles are displaced during drying~\cite{Anyfantakis2017} or small, undetected amounts of the nanoparticles contaminate the microparticle surfaces. In the suspension, these nanoparticles change the material response. 

The addition of the nanoparticles decreases the storage modulus was shown for rough particles where the dependence on the Hertzian contact force decreased with the roughness~\cite{Allard2022}. For the rough particle system, however, the LVE is extended with the rough particles due to interlocking of the asperities~\cite{Allard2022}. Such an effect is not present in the NP system due to the rotational and transnational freedom of the nanoparticles. These latter factors are particularly important in the compression experiments where the two main friction mechanisms are rolling and sliding~\cite{Tevet2011, Dai2016, Li2006}. The reduction in the Hertzian contact force combined with the reduction in friction allows the NP system to be more easily restructured and prevents the translation of the force from the upper plate to the lower.  

The combination of both a reduction in Hertzian contacts and promoted liquid exchange is hypothesized to be necessary to reduce network rigidity since a decrease in the coordination number and clustering coefficient was observed for the rough particle system~\cite{Allard2022} whereas the coordination number was increased in the present system. This hypothesis should be directly tested in compression experiments. The relative importance of frictional interactions (e.g. sliding versus rotation) should be tested since the joint effect of adhesion and friction has been shown to exacerbate shear thickening~\cite{Hsu2021} and be linked to arrest in depletion gels~\cite{vanderMeer2022}. Further, the influence of the nanoparticle properties (e.g. hydrophobicity and size ratios) should also be directly tested.  

These effects of nanoparticles in capillary suspensions are believed to be applicable to other systems, such as in wet granular materials where air can be seen as analogous to the bulk liquid in the present system. Thus, the nanoparticles are expected to act as flow aids and enhance microparticle sliding accompanied by the mitigated the liquid-air contact line pinning. Meanwhile, the nanoparticle induced thin films should be able to create inter-bridge connections which promote liquid redistribution. 

The present findings in this paper pose a challenge for future rheological measurements of particle networks. Even for samples without nanoparticles, a 30~$\mathrm{\mu}$m compression is able to result in an increase in the microparticle volume fraction of 4~vol\% and a coordination number of 0.2. This effect is compounded with the presence of nanoparticles. For example, wet granular systems are typically tested by either compressing sample between parallel plates and recording the normal force at the desired gap~\cite{Sweeney2017}, or the gap is set by the desired normal force~\cite{Badetti2018}. As shown in the present work, compression leads to restructuring and redistribution of the liquid bridges. Thus, the system is sensitive to the normal force, particularly when nanoparticles are present. Indeed, this work was motivated in part by the challenges observed in producing reliable rheological measurements. Thus, care should taken to ensure that the prepared sample structure is reproduced on the rheometer for wet granular systems and even in particle gels. That said, there are some applications where easy network restructuring while retaining a yield stress is important. One key area is in the 3D printing of pastes where nozzle clogging poses a challenge for both the creation of small feature sizes and in translating these into highly parallelized industrial processes.

\section*{Author Contributions}

\textbf{Lingyue Liu}: Conceptualization, Methodology, Software, Investigation, Writing – original draft. \textbf{Jens Allard}: Conceptualization, Methodology, Software, Writing – original draft. \textbf{Erin Koos}: Conceptualization, Methodology, Writing – review \& editing, Supervision.

\section*{Conflicts of interest}
There are no conflicts to declare.

\section*{Acknowledgements}
The authors would like to thank financial support from the European Union’s Horizon 2020 research and innovation programme under the Marie Sk\l{}odowska-Curie grant agreement No 955612
and International Fine Particle Research Institute (IFPRI).
\FloatBarrier


\begin{thebibliography}{10}
\expandafter\ifx\csname url\endcsname\relax
  \def\url#1{\texttt{#1}}\fi
\expandafter\ifx\csname urlprefix\endcsname\relax\def\urlprefix{URL }\fi
\expandafter\ifx\csname href\endcsname\relax
  \def\href#1#2{#2} \def\path#1{#1}\fi

\bibitem{Yang2017}
W.~Yang, Y.~Wu, X.~Pei, F.~Zhou, Q.~Xue, Contribution of surface chemistry to
  the shear thickening of silica nanoparticle suspensions, Langmuir 33~(4)
  (2017) 1037--1042.
\newblock \href {https://doi.org/10.1021/acs.langmuir.6b04060}
  {\path{doi:10.1021/acs.langmuir.6b04060}}.

\bibitem{Yang2005}
J.~Yang, A.~Sliva, A.~Banerjee, R.~N. Dave, R.~Pfeffer, Dry particle coating
  for improving the flowability of cohesive powders, Powder Technology 158~(1)
  (2005) 21--33.
\newblock \href {https://doi.org/10.1016/j.powtec.2005.04.032}
  {\path{doi:10.1016/j.powtec.2005.04.032}}.

\bibitem{Kim2022}
S.~S. Kim, C.~Castillo, M.~Sayedahmed, R.~N. Dav{\'e}, Reduced fine {API}
  agglomeration after dry coating for enhanced blend uniformity and
  processability of low drug loaded blends, Pharmaceutical Research 39~(12)
  (2022) 3155--3174.
\newblock \href {https://doi.org/10.1007/s11095-022-03343-6}
  {\path{doi:10.1007/s11095-022-03343-6}}.

\bibitem{Friedrich2010}
R.~B. Friedrich, M.~C. Fontana, M.~O. Bastos, A.~R. Pohlmann, S.~S. Guterres,
  R.~C.~R. Beck, Drying polymeric drug-loaded nanocapsules: {The} wet
  granulation process as a promising approach, Journal of Nanoscience and
  Nanotechnology 10~(1) (2010) 616--621.
\newblock \href {https://doi.org/10.1166/jnn.2010.1732}
  {\path{doi:10.1166/jnn.2010.1732}}.

\bibitem{Gondret1997}
P.~Gondret, L.~Petit, Dynamic viscosity of macroscopic suspensions of bimodal
  sized solid spheres, Journal of Rheology 41~(6) (1997) 1261--1274.
\newblock \href {https://doi.org/10.1122/1.550850}
  {\path{doi:10.1122/1.550850}}.

\bibitem{Probstein1994}
R.~F. Probstein, M.~Z. Sengun, T.~C. Tseng, Bimodal model of concentrated
  suspension viscosity for distributed particle sizes, Journal of Rheology
  38~(4) (1994) 811--829.
\newblock \href {https://doi.org/10.1122/1.550594}
  {\path{doi:10.1122/1.550594}}.

\bibitem{Chong1971}
J.~S. Chong, E.~B. Christiansen, A.~D. Baer, Rheology of concentrated
  suspensions, Journal of applied polymer science 15~(8) (1971) 2007--2021.
\newblock \href {https://doi.org/10.1002/app.1971.070150818}
  {\path{doi:10.1002/app.1971.070150818}}.

\bibitem{Greenwood1997}
R.~Greenwood, P.~F. Luckham, T.~Gregory, The effect of diameter ratio and
  volume ratio on the viscosity of bimodal suspensions of polymer latices,
  Journal of colloid and interface science 191~(1) (1997) 11--21.
\newblock \href {https://doi.org/10.1006/jcis.1997.4915}
  {\path{doi:10.1006/jcis.1997.4915}}.

\bibitem{McKee2012}
C.~T. McKee, J.~Y. Walz, Interaction forces between colloidal particles in a
  solution of like-charged, adsorbing nanoparticles, Journal of Colloid and
  Interface Science 365~(1) (2012) 72--80.
\newblock \href {https://doi.org/10.1016/j.jcis.2011.09.015}
  {\path{doi:10.1016/j.jcis.2011.09.015}}.

\bibitem{Kim2023}
S.~S. Kim, A.~Seetahal, N.~Amores, C.~Kossor, R.~N. Dav{\'e}, Impact of silica
  dry coprocessing with {API} and blend mixing time on blend flowability and
  drug content uniformity, Journal of Pharmaceutical Sciences (2023).
\newblock \href {https://doi.org/10.1016/j.xphs.2023.05.016}
  {\path{doi:10.1016/j.xphs.2023.05.016}}.

\bibitem{Koos2011}
E.~Koos, N.~Willenbacher, Capillary forces in suspension rheology, Science
  331~(6019) (2011) 897--900.
\newblock \href {https://doi.org/10.1126/science.1199243}
  {\path{doi:10.1126/science.1199243}}.

\bibitem{Koos2014}
E.~Koos, Capillary suspensions: {Particle} networks formed through the
  capillary force, Current Opinion in Colloid \& Interface Science 19~(6)
  (2014) 575--584.
\newblock \href {https://doi.org/10.1016/j.cocis.2014.10.004}
  {\path{doi:10.1016/j.cocis.2014.10.004}}.

\bibitem{Bossler2016}
F.~Bossler, E.~Koos, Structure of particle networks in capillary suspensions
  with wetting and nonwetting fluids, Langmuir 32~(6) (2016) 1489--1501.
\newblock \href {https://doi.org/10.1021/acs.langmuir.5b04246}
  {\path{doi:10.1021/acs.langmuir.5b04246}}.

\bibitem{Bindgen2020}
S.~Bindgen, F.~Bossler, J.~Allard, E.~Koos, Connecting particle clustering and
  rheology in attractive particle networks, Soft Matter 16~(36) (2020)
  8380--8393.
\newblock \href {https://doi.org/10.1039/D0SM00861C}
  {\path{doi:10.1039/D0SM00861C}}.

\bibitem{Dittmann2016}
J.~Dittmann, J.~Maurath, B.~Bitsch, N.~Willenbacher, Highly porous materials
  with unique mechanical properties from smart capillary suspensions, Advanced
  Materials 28~(8) (2016) 1689--1696.
\newblock \href {https://doi.org/10.1002/adma.201504910}
  {\path{doi:10.1002/adma.201504910}}.

\bibitem{Fischer2021a}
S.~B. Fischer, E.~Koos, Using an added liquid to suppress drying defects in
  hard particle coatings, Journal of Colloid and Interface Science 582 (2021)
  1231--1242.
\newblock \href {https://doi.org/10.1016/j.jcis.2020.08.055}
  {\path{doi:10.1016/j.jcis.2020.08.055}}.

\bibitem{Hoffmann2014}
S.~Hoffmann, E.~Koos, N.~Willenbacher, Using capillary bridges to tune
  stability and flow behavior of food suspensions, Food Hydrocolloids 40 (2014)
  44--52.
\newblock \href {https://doi.org/10.1016/j.foodhyd.2014.01.027}
  {\path{doi:10.1016/j.foodhyd.2014.01.027}}.

\bibitem{Dittmann2013}
J.~Dittmann, E.~Koos, N.~Willenbacher, Ceramic capillary suspensions: {Novel}
  processing route for macroporous ceramic materials, Journal of the American
  Ceramic Society 96~(2) (2013) 391--397.
\newblock \href {https://doi.org/10.1111/jace.12126}
  {\path{doi:10.1111/jace.12126}}.

\bibitem{Wollgarten2016}
S.~Wollgarten, C.~Yuce, E.~Koos, N.~Willenbacher, Tailoring flow behavior and
  texture of water based cocoa suspensions, Food Hydrocolloids 52 (2016)
  167--174.
\newblock \href {https://doi.org/10.1016/j.foodhyd.2015.06.010}
  {\path{doi:10.1016/j.foodhyd.2015.06.010}}.

\bibitem{Aksoy2023}
Y.~T. Aksoy, L.~Liu, M.~Abboud, M.~R. Vetrano, E.~Koos, Role of nanoparticles
  in nanofluid droplet impact on solid surfaces, Langmuir 39~(1) (2023) 12--19.
\newblock \href {https://doi.org/10.1021/acs.langmuir.2c02578}
  {\path{doi:10.1021/acs.langmuir.2c02578}}.

\bibitem{Park2019}
J.~Park, N.~Willenbacher, K.~H. Ahn, How the interaction between
  styrene-butadiene-rubber ({SBR}) binder and a secondary fluid affects the
  rheology, microstructure and adhesive properties of capillary-suspension-type
  graphite slurries used for {Li}-ion battery anodes, Colloids and Surfaces A:
  Physicochemical and Engineering Aspects 579 (2019) 123692.
\newblock \href {https://doi.org/10.1016/j.colsurfa.2019.123692}
  {\path{doi:10.1016/j.colsurfa.2019.123692}}.

\bibitem{Bossler2017}
F.~Bossler, L.~Weyrauch, R.~Schmidt, E.~Koos, Influence of mixing conditions on
  the rheological properties and structure of capillary suspensions, Colloids
  and Surfaces A: Physicochemical and Engineering Aspects 518 (2017) 85--97.
\newblock \href {https://doi.org/10.1016/j.colsurfa.2017.01.026}
  {\path{doi:10.1016/j.colsurfa.2017.01.026}}.

\bibitem{Weis2020}
M.~Wei{\ss}, P.~S\"{a}lzler, N.~Willenbacher, E.~Koos, {3D}-{Printed}
  lightweight ceramics using capillary suspensions with incorporated
  nanoparticles, Journal of the European Ceramic Society 40~(8) (2020)
  3140--3147.
\newblock \href {https://doi.org/10.1016/j.jeurceramsoc.2020.02.055}
  {\path{doi:10.1016/j.jeurceramsoc.2020.02.055}}.

\bibitem{Bossler2018}
F.~Bossler, J.~Maurath, K.~Dyhr, N.~Willenbacher, E.~Koos, Fractal approaches
  to characterize the structure of capillary suspensions using rheology and
  confocal microscopy, Journal of Rheology 62~(1) (2018) 183--196.
\newblock \href {https://doi.org/10.1122/1.4997889}
  {\path{doi:10.1122/1.4997889}}.

\bibitem{Allard2022}
J.~Allard, S.~Burgers, M.~C. {Rodr\'{i}guez Gonz\'{a}lez}, Y.~Zhu, S.~{De
  Feyter}, E.~Koos, Effects of particle roughness on the rheology and structure
  of capillary suspensions, Colloids and Surfaces {A}: Physicochemical and
  Engineering Aspects 648 (2022) 129224.
\newblock \href {https://doi.org/10.1016/j.colsurfa.2022.129224}
  {\path{doi:10.1016/j.colsurfa.2022.129224}}.

\bibitem{Tevet2011}
O.~Tevet, P.~Von-Huth, R.~Popovitz-Biro, R.~Rosentsveig, H.~D. Wagner,
  R.~Tenne, Friction mechanism of individual multilayered nanoparticles,
  Proceedings of the National Academy of Sciences 108~(50) (2011) 19901--19906.
\newblock \href {https://doi.org/10.1073/pnas.1106553108}
  {\path{doi:10.1073/pnas.1106553108}}.

\bibitem{Dai2016}
W.~Dai, B.~Kheireddin, H.~Gao, H.~Liang, Roles of nanoparticles in oil
  lubrication, Tribology International 102 (2016) 88--98.
\newblock \href {https://doi.org/10.1016/j.triboint.2016.05.020}
  {\path{doi:10.1016/j.triboint.2016.05.020}}.

\bibitem{Halsey1998}
T.~C. Halsey, A.~J. Levine, How {Sandcastles} {Fall}, Physical Review Letters
  80~(14) (1998) 3141--3144.
\newblock \href {https://doi.org/10.1103/PhysRevLett.80.3141}
  {\path{doi:10.1103/PhysRevLett.80.3141}}.

\bibitem{Suratwala2003}
T.~I. Suratwala, M.~L. Hanna, E.~L. Miller, P.~K. Whitman, I.~M. Thomas, P.~R.
  Ehrmann, R.~S. Maxwell, A.~K. Burnham, Surface chemistry and trimethylsilyl
  functionalization of {Stöber} silica sols, Journal of Non-Crystalline Solids
  316~(2) (2003) 349--363.
\newblock \href {https://doi.org/10.1016/S0022-3093(02)01629-0}
  {\path{doi:10.1016/S0022-3093(02)01629-0}}.

\bibitem{MCR702}
J.~L\"{a}uger, in: An enhanced rotational rheometer system with two motors,
  2015, pp. 55--61.

\bibitem{Weeks2000}
E.~R. Weeks, J.~C. Crocker, A.~C. Levitt, A.~Schofield, D.~A. Weitz,
  Three-dimensional direct imaging of structural relaxation near the colloidal
  glass transition, Science 287~(5453) (2000) 627--631.
\newblock \href {https://doi.org/10.1126/science.287.5453.627}
  {\path{doi:10.1126/science.287.5453.627}}.

\bibitem{Scheel2008}
M.~Scheel, R.~Seemann, M.~Brinkmann, M.~{Di Michiel}, A.~Sheppard,
  B.~Breidenbach, S.~Herminghaus, Morphological clues to wet granular pile
  stability, Nature Materials 7~(3) (2008) 189--193.
\newblock \href {https://doi.org/10.1038/nmat2117}
  {\path{doi:10.1038/nmat2117}}.

\bibitem{Hamamoto2015}
M.~Hamamoto, M.~Katsura, N.~Nishiyama, R.~Tononue, S.~Nakashima, Transmission
  {IR} micro-spectroscopy of interfacial water between colloidal silica
  particles, e-Journal of Surface Science and Nanotechnology 13 (2015)
  301--306.
\newblock \href {https://doi.org/10.1380/ejssnt.2015.301}
  {\path{doi:10.1380/ejssnt.2015.301}}.

\bibitem{Sulpizi2012}
M.~Sulpizi, M.-P. Gaigeot, M.~Sprik, The silica–water interface: {How} the
  silanols determine the surface acidity and modulate the water properties,
  Journal of Chemical Theory and Computation 8~(3) (2012) 1037--1047.
\newblock \href {https://doi.org/10.1021/ct2007154}
  {\path{doi:10.1021/ct2007154}}.

\bibitem{Lu2014}
G.~Lu, Y.-Y. Duan, X.-D. Wang, Surface tension, viscosity, and rheology of
  water-based nanofluids: a microscopic interpretation on the molecular level,
  Journal of Nanoparticle Research 16~(9) (2014) 2564.
\newblock \href {https://doi.org/10.1007/s11051-014-2564-2}
  {\path{doi:10.1007/s11051-014-2564-2}}.

\bibitem{Hamaker1937}
H.~C. Hamaker, The {London}—van der {Waals} attraction between spherical
  particles, Physica 4~(10) (1937) 1058--1072.
\newblock \href {https://doi.org/10.1016/S0031-8914(37)80203-7}
  {\path{doi:10.1016/S0031-8914(37)80203-7}}.

\bibitem{Alvo2010}
S.~Alvo, P.~Lambert, M.~Gauthier, S.~Régnier, A van der {Waals} force-based
  adhesion model for micromanipulation., Journal of Adhesion Science and
  Technology 24~(15-16) (2010) 2415--2428.
\newblock \href {https://doi.org/10.1163/0169942410X508334}
  {\path{doi:10.1163/0169942410X508334}}.

\bibitem{Bresme2007}
F.~Bresme, M.~Oettel, Nanoparticles at fluid interfaces, Journal of Physics.
  Condensed Matter: An Institute of Physics Journal 19~(41) (2007) 413101.
\newblock \href {https://doi.org/10.1088/0953-8984/19/41/413101}
  {\path{doi:10.1088/0953-8984/19/41/413101}}.

\bibitem{Kralchevsky1994}
P.~A. Kralchevsky, K.~Nagayama, Capillary forces between colloidal particles,
  Langmuir 10~(1) (1994) 23--36.
\newblock \href {https://doi.org/10.1021/la00013a004}
  {\path{doi:10.1021/la00013a004}}.

\bibitem{Ortiz2013}
D.~Ortiz-Young, H.-C. Chiu, S.~Kim, K.~Voïtchovsky, E.~Riedo, The interplay
  between apparent viscosity and wettability in nanoconfined water, Nature
  Communications 4~(1) (2013) 2482.
\newblock \href {https://doi.org/10.1038/ncomms3482}
  {\path{doi:10.1038/ncomms3482}}.

\bibitem{Anyfantakis2017}
M.~Anyfantakis, D.~Baigl, B.~P. Binks, Evaporation of drops containing silica
  nanoparticles of varying hydrophobicities: {Exploiting} particle-particle
  interactions for additive-free tunable deposit morphology, Langmuir 33~(20)
  (2017) 5025--5036.
\newblock \href {https://doi.org/10.1021/acs.langmuir.7b00807}
  {\path{doi:10.1021/acs.langmuir.7b00807}}.

\bibitem{Donley2020}
G.~J. Donley, P.~K. Singh, A.~Shetty, S.~A. Rogers, Elucidating the {$G''$}
  overshoot in soft materials with a yield transition via a time-resolved
  experimental strain decomposition, Proceedings of the National Academy of
  Sciences 117~(36) (2020) 21945--21952.
\newblock \href {https://doi.org/10.1073/pnas.2003869117}
  {\path{doi:10.1073/pnas.2003869117}}.

\bibitem{Natalia2022}
I.~Natalia, R.~H. Ewoldt, E.~Koos, Particle contact dynamics as the origin for
  noninteger power expansion rheology in attractive suspension networks,
  Journal of Rheology 66~(1) (2022) 17--30.
\newblock \href {https://doi.org/10.1122/8.0000289}
  {\path{doi:10.1122/8.0000289}}.

\bibitem{Donley2019}
G.~J. Donley, J.~R. de~Bruyn, G.~H. McKinley, S.~A. Rogers, Time-resolved
  dynamics of the yielding transition in soft materials, Journal of
  {Non-Newtonian} Fluid Mechanics 264 (2019) 117--134.
\newblock \href {https://doi.org/10.1016/j.jnnfm.2018.10.003}
  {\path{doi:10.1016/j.jnnfm.2018.10.003}}.

\bibitem{Timoshenko1959}
S.~Timoshenko, S.~Woinowsky-Krieger, Theory of plates and shells, 2nd Edition,
  Engineering societies monographs, McGraw-Hill, New York, 1959.

\bibitem{Young2012}
W.~C. Young, R.~G. Budynas, A.~M. Sadegh, Roark's Formulas for Stress and
  Strain, 8th Edition, McGraw-Hill Education, New York, 2012.

\bibitem{Bindgen2022}
S.~Bindgen, J.~Allard, E.~Koos, The behavior of capillary suspensions at
  diverse length scales: from single capillary bridges to bulk, Current Opinion
  in Colloid \& Interface Science 58 (2022) 101557.
\newblock \href {https://doi.org/10.1016/j.cocis.2021.101557}
  {\path{doi:10.1016/j.cocis.2021.101557}}.

\bibitem{Brodu2015}
N.~Brodu, J.~A. Dijksman, R.~P. Behringer, Spanning the scales of granular
  materials through microscopic force imaging, Nature Communications 6 (2015)
  6361.
\newblock \href {https://doi.org/10.1038/ncomms7361}
  {\path{doi:10.1038/ncomms7361}}.

\bibitem{Nguyen2019}
H.~T. Nguyen, A.~L. Graham, P.~H. Koenig, L.~D. Gelb, Computer simulations of
  colloidal gels: how hindered particle rotation affects structure and
  rheology, Soft Matter 16~(1) (2019) 256--269.
\newblock \href {https://doi.org/10.1039/C9SM01755K}
  {\path{doi:10.1039/C9SM01755K}}.

\bibitem{Wang2017}
J.-P. Wang, E.~Gallo, B.~Fran\c{c}ois, F.~Gabrieli, P.~Lambert, Capillary force
  and rupture of funicular liquid bridges between three spherical bodies,
  Powder Technology 305 (2017) 89--98.
\newblock \href {https://doi.org/10.1016/j.powtec.2016.09.060}
  {\path{doi:10.1016/j.powtec.2016.09.060}}.

\bibitem{Gao2018}
N.~Gao, F.~Geyer, D.~Pilat, S.~Wooh, D.~Vollmer, H.-J. Butt, R.~Berger, How
  drops start sliding over solid surfaces, Nature Physics 14 (2018) 191–196.
\newblock \href {https://doi.org/10.1038/nphys4305}
  {\path{doi:10.1038/nphys4305}}.

\bibitem{Song2021}
Q.~Song, K.~Liu, W.~Sun, R.~Chen, J.~Ji, Y.~Jiao, T.~Gao, J.~Ye, Lateral and
  normal capillary force evolution of a reciprocating liquid bridge, Langmuir
  37~(40) (2021) 11737--11749.
\newblock \href {https://doi.org/10.1021/acs.langmuir.1c01635}
  {\path{doi:10.1021/acs.langmuir.1c01635}}.

\bibitem{Pilat2012}
D.~W. Pilat, P.~Papadopoulos, D.~Sch\"{a}ffel, D.~Vollmer, R.~Berger, H.-J.
  Butt, Dynamic measurement of the force required to move a liquid drop on a
  solid surface, Langmuir 28~(49) (2012) 16812--16820.
\newblock \href {https://doi.org/10.1021/la3041067}
  {\path{doi:10.1021/la3041067}}.

\bibitem{Hntsberger1981}
J.~R. Hntsberger, Surface energy, wetting and adhesion, The Journal of Adhesion
  12~(1) (1981) 3--12.
\newblock \href {https://doi.org/10.1080/00218468108071184}
  {\path{doi:10.1080/00218468108071184}}.

\bibitem{Wang2017b}
J.-P. Wang, Force transmission modes of non-cohesive and cohesive materials at
  the critical state, Materials 10~(9) (2017) 1014.
\newblock \href {https://doi.org/10.3390/ma10091014}
  {\path{doi:10.3390/ma10091014}}.

\bibitem{Ewoldt2013}
R.~H. Ewoldt, N.~A. Bharadwaj, Low-dimensional intrinsic material functions for
  nonlinear viscoelasticity, Rheologica Acta 52~(3) (2013) 201--219.
\newblock \href {https://doi.org/10.1007/s00397-013-0686-6}
  {\path{doi:10.1007/s00397-013-0686-6}}.

\bibitem{Li2006}
X.~Li, Z.~Cao, Z.~Zhang, H.~Dang, Surface-modification in situ of
  nano-{SiO$_2$} and its structure and tribological properties, Applied Surface
  Science 252~(22) (2006) 7856--7861.
\newblock \href {https://doi.org/10.1016/j.apsusc.2005.09.068}
  {\path{doi:10.1016/j.apsusc.2005.09.068}}.

\bibitem{Hsu2021}
C.-P. Hsu, J.~Mandal, S.~N. Ramakrishna, N.~D. Spencer, L.~Isa, Exploring the
  roles of roughness, friction and adhesion in discontinuous shear thickening
  by means of thermo-responsive particles, Nature communications 12~(1) (2021)
  1477.
\newblock \href {https://doi.org/10.1038/s41467-021-21580-y}
  {\path{doi:10.1038/s41467-021-21580-y}}.

\bibitem{vanderMeer2022}
B.~{van der Meer}, T.~Yanagishima, R.~Dullens, Emergence of interparticle
  friction in attractive colloidal matter, arXiv preprint arXiv:2209.12703
  (2022).

\bibitem{Sweeney2017}
M.~Sweeney, L.~L. Campbell, J.~Hanson, M.~L. Pantoya, G.~F. Christopher,
  Characterizing the feasibility of processing wet granular materials to
  improve rheology for {3D} printing, Journal of Materials Science 52~(22)
  (2017) 13040--13053.
\newblock \href {https://doi.org/10.1007/s10853-017-1404-z}
  {\path{doi:10.1007/s10853-017-1404-z}}.

\bibitem{Badetti2018}
M.~Badetti, A.~Fall, D.~Hautemayou, F.~Chevoir, P.~Aimedieu, S.~Rodts, J.-N.
  Roux, Rheology and microstructure of unsaturated wet granular materials:
  {Experiments} and simulations, Journal of Rheology 62~(5) (2018) 1175--1186.
\newblock \href {https://doi.org/10.1122/1.5026979}
  {\path{doi:10.1122/1.5026979}}.

\end{thebibliography}



\section*{Supplementary material}

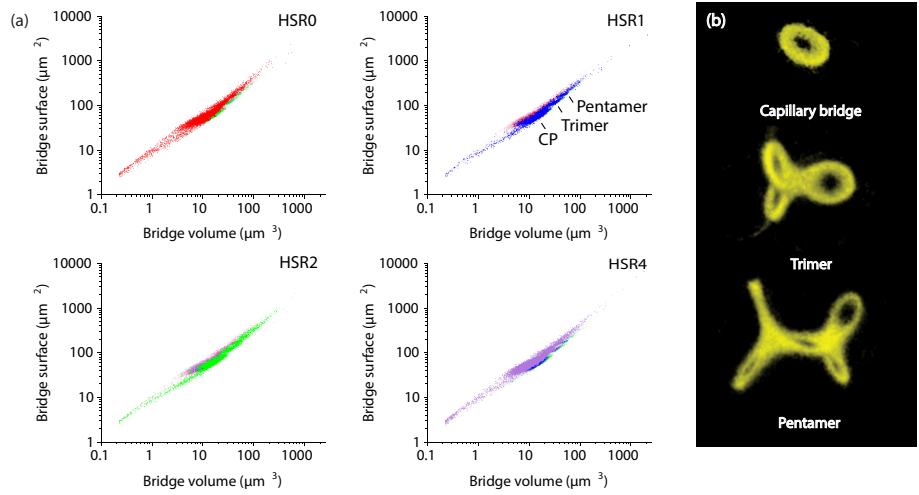

Figure S1: (a) Cumulative plot of secondary liquid surface area versus the volume, the HSR of interest is highlighted while the rest are faded in the background. (b) 3D view of different bridge types, capillary bridge (CP), trimer and pentamer, represented in (a).

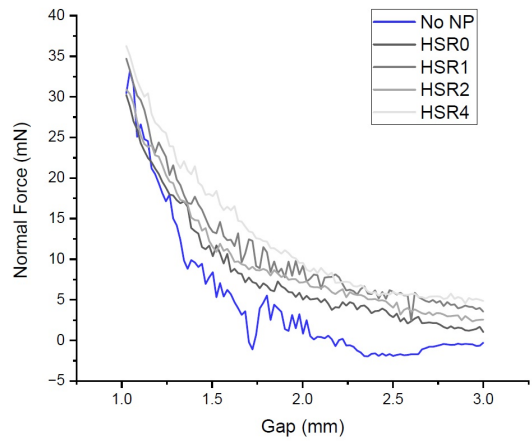

Figure S2: Normal force changes of samples without and with nanoparticles of different hydrophobicities versus rheometer gap height. All samples have similar final normal forces $F_N = 33 \pm 3$ mN.

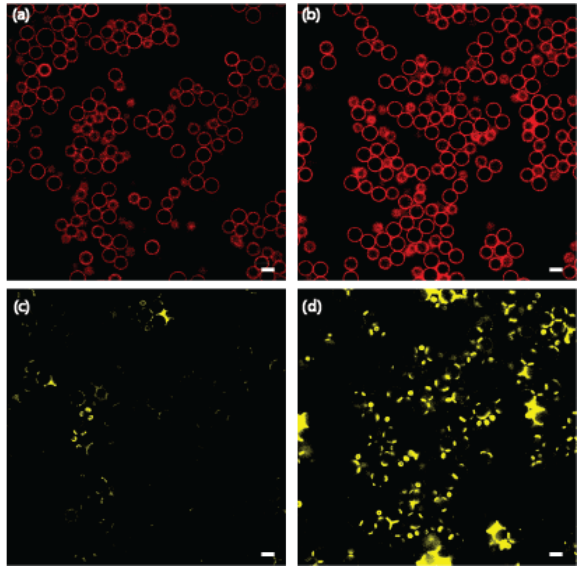

Figure S3: Confocal microscopical images of capillary suspensions with and without nanoparticles before compression. Red are the microparticles while the yellow represents the secondary liquid. (a) and (c) are the sample without nanoparticles, (b) and (d) are the samples with HSR0 nanoparticles.

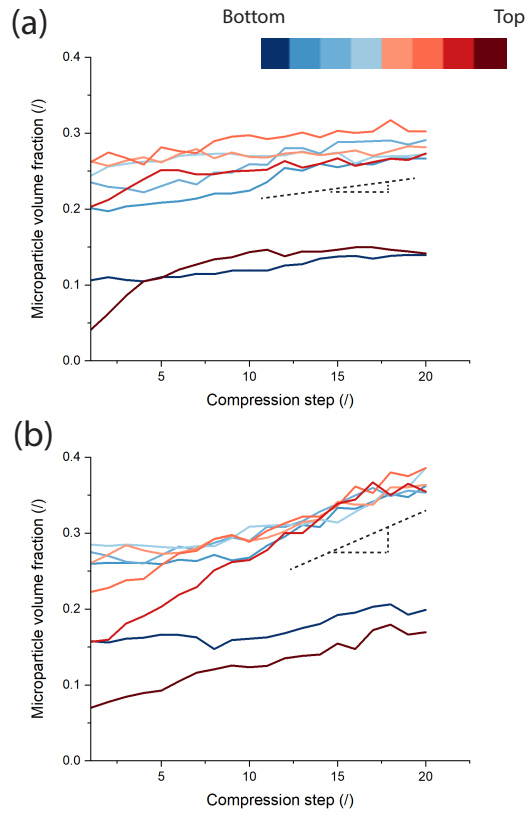

Figure S4: Development of microparticle volume fraction per layer (>10 $\mu$m) during compression for samples (a) without nanoparticles, and (b) with HSR0 nanoparticles. The color gradient from dark blue to red represents the layers from bottom to top. The layer thickness shrinks proportionally with the decrease in sample height. The dashed lines are the general slopes.

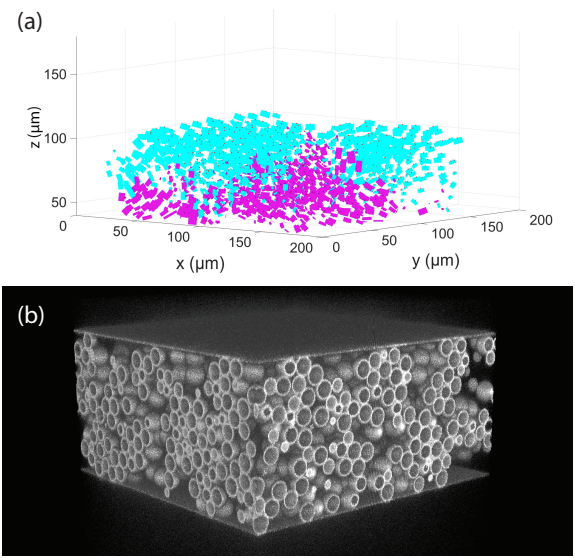

Figure S5: Particle movement of sample with HSR0 nanoparticles during homogeneous compression demonstrated in (a) the relative movement plot, cyan trajectories represent more vertical movements downwards, magenta trajectories represent the microparticles that resist compression, the line widths are proportional to their individual moving distances, red dots are the starting point of microparticles (d) 3d imaging of microparticle position at the beginning stage of the compression.

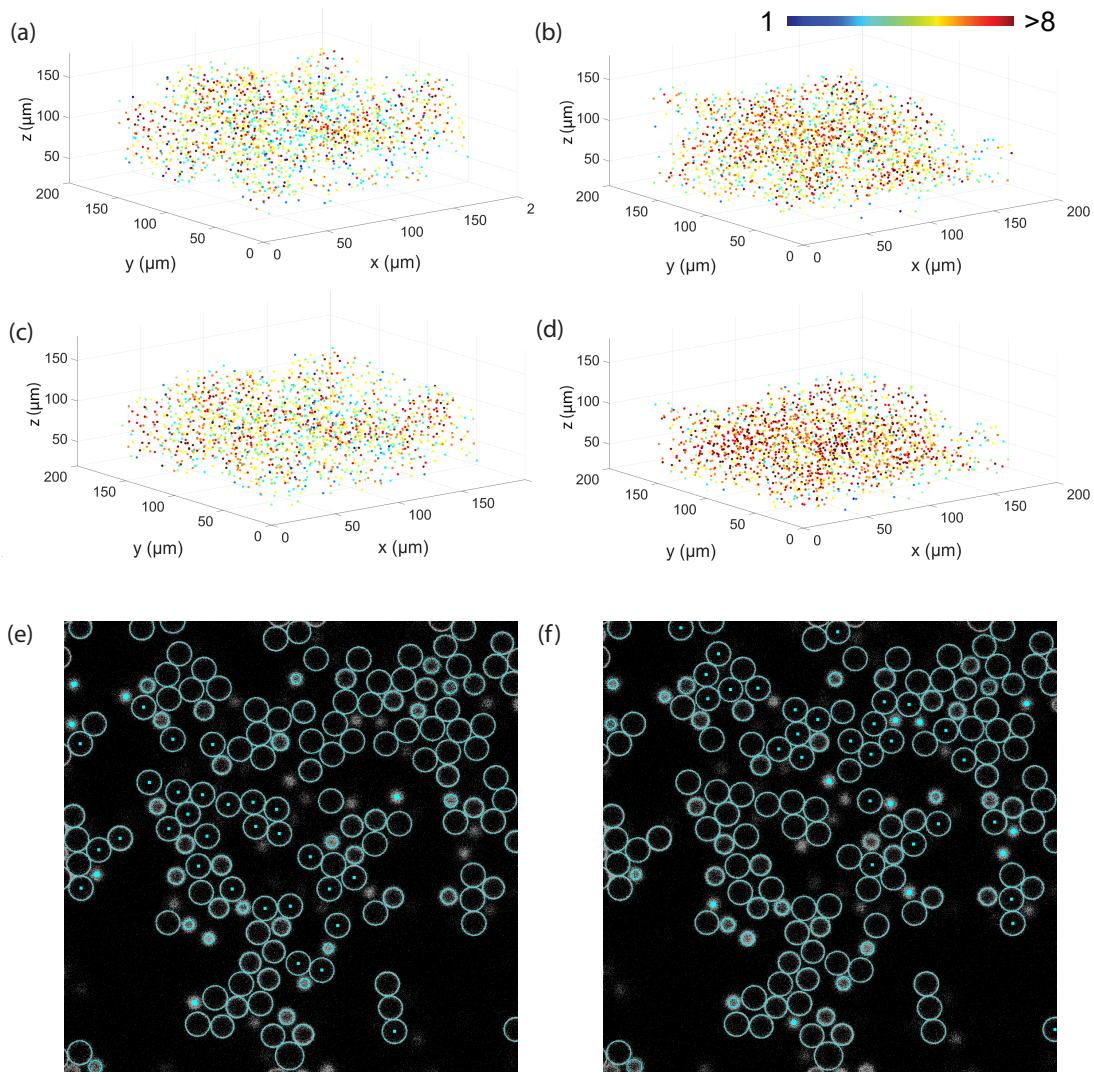

Figure S6: 3D representation of the capillary suspension networks without (a and c) and with (b and d) nanoparticles, pre and post compression respectively, colors are based on the individual coordination number. From dark blue to dark red, the coordination number increases. The superimposed original micrographs and location detected micrographs are as demonstrated in (e) and (f). The original images are presented in white, the sphere's center and outlines are presented in cyan. These two frames are 0.33 $\mu$m apart from each other.

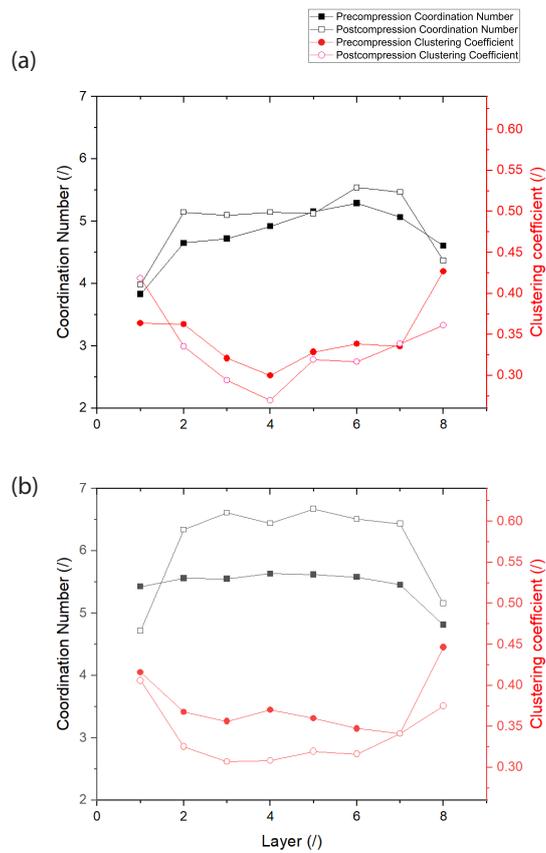

Figure S7: Development of coordination number and clustering coefficient per layer before and after compression for samples (a) without nanoparticles and (b) with HSR0 nanoparticles. The layer thickness shrinks proportionally with the decrease in sample height. The smaller the layer numbers, the closer they are to the top glass.

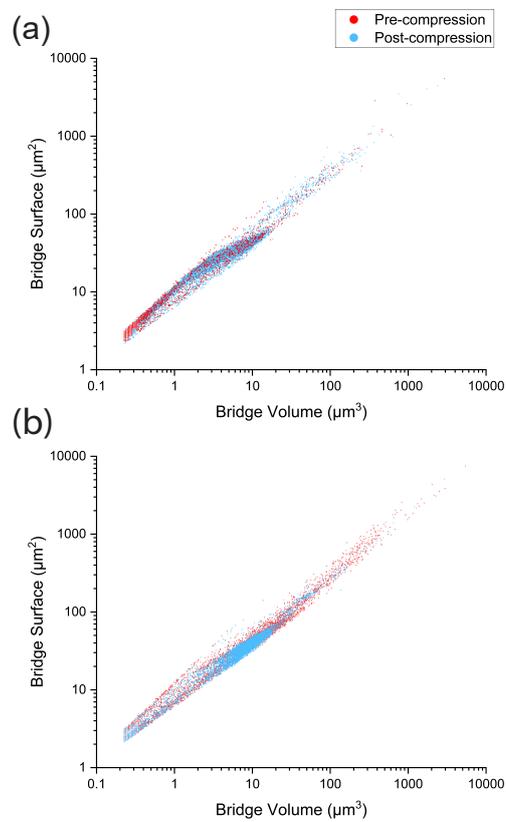

Figure S8: Cumulative plots of secondary liquid surface area versus the volume before and after compression of sample (a) without nanoparticles, and (b) with HSR0 nanoparticles.



\end{document}